\documentclass[review]{elsarticle}
\usepackage[utf8]{inputenc}
\usepackage{amsmath} 
\usepackage{caption}
\usepackage{float}
\usepackage{graphicx}
\usepackage{color}
\usepackage{subcaption}
\usepackage{ulem}

\usepackage{lineno,hyperref}

\begin{document}

\begin{frontmatter}

\title{Selection Principles for Gaia}

\author{ Rudy ~Arthur${}^{1*}$, Arwen ~Nicholson${}^2$ }
\address{1. Department of Computer Science, University of Exeter, North Park Road,Exeter,UK,EX4 4RN,  \texttt{* R.Arthur@exeter.ac.uk}}
\address{2. Department of Physics, University of Exeter, North Park Road,Exeter, UK, EX4 4QL}

\begin{abstract}
The Gaia hypothesis considers the life-environment coupled system as a single entity that acts to regulate and maintain habitable conditions on Earth.
In this paper we discuss three mechanisms which could potentially lead to Gaia: Selection by Survival, Sequential Selection and Entropic Hierarchy. We use the Tangled Nature Model of co-evolution as a common framework for investigating all three, using an extended version of the standard model to elaborate on Gaia as an example of an entropic hierarchy. This idea, which combines sequential selection together with a reservoir of diversity that acts as a `memory', implies a tendency towards growth and increasing resilience of the Gaian system over time. We then discuss how Gaian memory could be realised in practice via the microbial seed bank, climate refugia and lateral gene transfer and conclude by discussing testable implications of an entropic hierarchy for the study of Earth history and the search for life in the universe. This paper adds to the existing taxonomy of Gaia hypotheses to suggest an ``Entropic Gaia" where we argue that increasing biomass, complexity and enhanced habitability over time is a statistically likely feature of a co-evolving system.
\end{abstract}


\end{frontmatter}

\section{Introduction}

The Gaia hypothesis postulates that life interacts with the Earth to form a self regulating system \cite{Lovelock:1974,Lovelock:2000,Lenton:1998}. To date, most models of Gaia have focused on showing how communities of organisms can spontaneously generate favourable environmental conditions or maintain them in the face of external perturbation \cite{Schneider:2013,Schwartzman:1989,Schwartzman:1991,Volk:1998,Walker:1981, Watson:1983, Wood:2008, Downing:1999, McDonaldGibson:2008,Williams:2007, Williams:2010, Nicholson:2017,Dyke:2013,Nicholson:2018}. There has been a particular focus on feedback mechanisms for maintaining temperature or chemical concentration within a particular habitable range, and several plausible ideas for how this might occur in practice have been proposed, for example niche construction \cite{Odling:2013} and rein-control \cite{Harvey:2004}. 

Recently there has been much discussion in the literature about selection principles \cite{Doolittle:2014,Bouchard:2014,Toman:2017,Nicholson:2018a}. Most of this work concerns, to use Dolittle's terminology \cite{Doolittle:2014}, \textit{selection by survival}. Distinct from Darwinian selection, which requires variation, inheritance and competition, selection by survival simply requires persistence. Differential survival among a variable population will inevitably lead, after a long time, to a population of individuals with persistence enhancing traits. This is more or less a tautology: the things which survive have properties which enable them to survive. Nevertheless, this principle operating in conjunction with natural selection over long timescales could explain certain biological facts e.g. the appearance of some evolutionary pre-adaptations \cite{Toman:2017}. 

A similar idea is \textit{sequential selection} \cite{Nicholson:2018b, Betts:2007}. Here we imagine life co-evolving with the environment. If life has a detrimental effect on its environment (e.g. resource depletion) a critical point will be reached and the system will collapse. After the collapse, a new life-environment system will emerge and co-evolve until it collapses. Similar to selection by survival, the systems we see persisting are the ones which have properties that allow persistence. Sequential selection can be viewed as selection by survival operating by repeated, sequential, trials, rather than simultaneously across a large population, as in selection by survival.

These selection principles would imply that we see Gaia for essentially probabilistic reasons \cite{Lenton:2003, Lenton:2011}. Gaia is realised in practice via feedback loops and mechanisms such as niche construction \cite{Odling:2013} and rein control \cite{Harvey:2004}. According to sequential selection and selection by survival the reason we observe these mechanisms is that they are persistence enhancing and we, as the observers, could only have evolved during a persistent state with mechanisms like these in effect.

Recent work on coupled species-environment models \cite{Arthur:2017} implies that there may be more than differential survival at work. This work studied systems which undergo sequential selection and demonstrate a spontaneous tendency towards increasing biomass, diversity and stability with time, while maintaining positive species-environment interactions. This implies a cumulative process where persistence enhancing system characteristics are preferentially selected. Therefore another effect, over and above differential survival, is operating in these systems.

Sequential selection and selection by survival imply that Gaia is not a fundamental feature of the coupled life-environment systems, but is a kind of observer effect or anthropic principle \cite{Smolin:2004}. In this work we will argue that Gaia - defined as the tendency of life towards improving its environment, diversity, total biomass and stability - is in fact a statistically likely outcome of a coupled life-environment system with `memory'. This paper will use the same modelling framework to simulate three selection principles:  selection by survival, sequential selection  and entropic hierarchy. The aim is  to show that, while the first two  lead  to  Gaia  by  differential  survival, the latter makes Gaia a statistically likely outcome for a co-evolutionary system. Differential survival implies Gaian systems are lucky accidents among a large number of failures and false starts, while the entropic hierarchy is a kind of ratcheting mechanism which any co-evolving system can undergo. This means we can understand the emergence of Gaia on Earth without the need for anthropic arguments. This has implications not only for our understanding of earth history \cite{Lenton:2011}, but also for exo-planet research \cite{Goldblatt:2015, Lenardic:2016} and issues such as the Fermi paradox \cite{Watson:2004}.

To situate this paper within the Gaia literature we note that the Gaia hypothesis has been defined in numerous ways over the decades, and a taxonomy of these was suggested by Kirchner \cite{Kirchner:1989}, based on quotes on Gaia from papers and books by Lovelock and Margulis, as follows:
\begin{description}
\item[Coevolutionary Gaia] Life influences its abiotic environment, and the environment in turn influences life.
\item[Homeostatic Gaia] Life influences the world in a way that leads to stability due to the dominant links between life and the abiotic world being negative feedback loops. This hypothesis is further broken down into:
\begin{description}
\item[Lucky Gaia] The Earth has homeostatic properties largely by luck \cite{Watson:2004}.
\item[Probable Gaia] The probability for a life-planet coupled system to develop homeostatic properties is greater than the probability to evolve non-homeostasis \cite{Lenton:2003}. 
\end{description}
\item[Geophysiological Gaia] The biosphere can be described as a single organism, which can exhibit both homeostatic and unstable behaviour, like other organisms.
\item[Optimising Gaia] Life interacts with its physical environment in such a way that it maintains optimum conditions for life at all times.
\end{description}
We suggest that this taxonomy of Gaia hypotheses can be extended. Our ecosystem models suggest the following hypothesis:\\ \newline
\textbf{Entropic Gaia} Co-evolutionary systems evolve in the direction of increasing information entropy. This direction corresponds to greater biomass, species diversity and life-enhancing abiotic interactions.

This paper is our argument for Entropic Gaia which we construct as follows. In section \ref{sec:statmech} we discuss how information entropy can be a useful concept in ecology and tie this to the familiar idea of a fitness landscape. In section \ref{sec:tnm} and \ref{sec:modeldes} we describe the Tangled Nature Model (TNM) \cite{Christensen:2002}, motivating it by linking it to the generalised Lotka-Volterra equations. The TNM will be the main framework we use to explore the ideas of selection by survival in section \ref{sec:sbs} and sequential selection in section \ref{sec:ss}. We study the full model in section \ref{sec:tnmsim}, discussing how the model dynamics lead to distinctly Gaian features - namely increased biomass, diversity, stability and favourable life-environment interaction. Motivated by the TNM results, in section \ref{sec:tnment} we explain the idea of an entropic hierarchy and posit it as a mechanism for `generating' Gaia. For this mechanism to work in the real earth system we need some kind of store of diversity or memory; we discuss three plausible realisations of this in section \ref{sec:divres}. Finally in section \ref{sec:conc} we summarise our results and suggest some testable implications of the entropic hierarchy idea.

\section{Ecology and Entropy}\label{sec:statmech}

Wright's idea of the fitness landscape \cite{Wright:1932} is a familiar language for mathematical biologists. This idea has been quite influential for reasoning about and explicitly modelling evolution as an optimisation process e.g. \cite{Kauffman:1989}. This metaphor constructs a gene space with some function called fitness defined at every point. A particular individual is a point in this space and evolution can be thought of as an optimisation algorithm that seeks to maximise  fitness by finding peaks on the landscape. 

In this paper we will consider collections of interacting species and use the language of statistical thermodynamics and information entropy to characterise the behaviour of these systems. As this framework is new for models of Gaia and may be unfamiliar to researchers in the field, this section serves as a brief introduction to the language and serves to establish a `landscape' metaphor appropriate for studying co-evolution. Readers interested in a more thorough treatment of these ideas are referred to Jaynes \cite{Jaynes:1957} and Ulanowicz \cite{Ulanowicz:2001}. 

\subsection{Micro and Macro-States}

For many systems in nature their observed macro properties e.g. temperature, pH or pressure are due to their micro properties e.g. the position, velocity and type of their constituent particles. Each macro-state can be realised by many possible micro-states while every micro-state corresponds to a particular macro-state. 
From a Gaian perspective the macro-state is the global ecosystem, characterised by certain bulk properties like total biomass, average temperature, oceanic salinity, amount of $CO_2$ in the atmosphere etc. and the micro-state is the particular combination of organisms and geology that gives rise to it. Each possible configuration of individual organisms gives rise to a particular global environment, but each global environment can be realised by many different collections of organisms. 

In statistical mechanics systems move from one micro-state to another via a random walk. The system will therefore tend to be in the macro-state which is realised by the largest number of micro-states. The \textbf{entropy} of a macro-state is the log of the number of micro-states which can realise it, hence the entropy of a system increases with time, as the system moves towards the most probable macro-state, which is realised by the largest number of micro-states.

A system of particles will move through the space of all possible micro-states, called  \textbf{configuration-space}, and settle in an area corresponding to the most probable macro-state, which we recognise as the system's equilibrium. However, for species which can mutate it is possible for new species to arise which can disrupt the equilibrium by reshaping the fitness landscapes of all species. For example, a novel virus which affects a keystone species. After the disruption, the system will find a new equilibrium which will not be the same as the last one. This leads to observations of \textit{punctuated equilibrium} \cite{Gould:1972,Bak:1993} and we speak of \textit{quasi}-equilibrium.

We note a point of possible confusion between two related but different notions of entropy. Life has low \textit{thermodynamic entropy} which it maintains by exchanging heat and matter with the environment. This form of entropy concerns the physical and chemical processes each individual organism is involved in. In this paper we are talking exclusively about \textit{information entropy}, also known as Shannon entropy \cite{Shannon:1948}. This is an abstract quantity which is related to the volume of configuration space that the ecosystem occupies. When we talk about `entropy increasing' we mean the information entropy, the thermodynamic entropy of any particular organism is not affected.

\subsection{Ecological Fitness Landscapes}\label{sec:fl}

In a multi-species system, a change in the genotype of one species to raise its fitness has an effect on the fitness of all the other species with which it interacts and \textit{vice versa}. The static landscape picture is thus not appropriate for an interacting, multi-species system. The metaphor of a fitness seascape \cite{Mustonen:2009} has been used to describe the system of interacting and fluctuating fitness landscapes of each species.

As argued in \cite{Arthur:2017a} by `moving up a level' from the individual to the ecosystem, we can restore the static landscape metaphor. Assume we have a number of species, $D$, with species labelled $i$ each having population $N_i$. Define the total population as $N = \sum_i^D N_i$. The usual fitness landscape is a function which assigns some number $f_i$ to each genotype $i$,  where higher values of $f_i$ are associated with an increased likelihood to reproduce. The issue identified above is that the fitness of a genotype depends on the environment and the other species that are present. A fox may be very fit in an environment with many rabbits as prey, but not in an environment with one or fewer rabbits.

The genotype fitness $f_i$ is thus a function not only of genotype, $i$, but of the type and abundance of every species in the system. Every evolutionary innovation or population shift has cascading effects on the fitness of every individual in the system. To deal with this, define a function
\begin{equation}\label{eqn:landscape}
    F(N_1, N_2, \ldots, N_D) = \sum_i N_i f_i
\end{equation}
which is a sum of the population weighted fitness of every species. $F$ is a kind of `ecosystem fitness', as shown in section \ref{sec:tnm} it is the sum of all the growth rates in a generalised Lotka-Volterra model. $F$ is a function of the population of every possible species and larger values of $F$ imply more reproduction is happening in the system. 

The `configuration space' discussed above can be thought of as the space defined by the variables $N_1, N_2, \ldots, N_D$, where $N_i \geq 0$ and the `landscape' is defined by the function $F(N_1, N_2, \ldots, N_D)$. An ecosystem is a point in this space and as the species grow and evolve the point moves around. Ecosystems typically do not compete, so there is no Darwinian optimisation on this landscape. Instead the point in configuration space undergoes a random walk. Configuration space is the stage for all of our discussions about selection principles. As we will discuss in the following, selection by survival and sequential selection imply a completely random search in configuration space, while the entropic mechanism we discuss implies a weak `force' driving systems uphill.

\section{Tangled Nature Model}\label{sec:tnm}

The Tangled Nature Model (TNM) of Christensen et. al. \cite{Christensen:2002} is an agent based model designed to explore the idea of co-evolution. In this section we will argue, building on the ideas in \cite{Arthur:2017}, that the TNM is not simply a toy model. We will show that the average behaviour of the TNM describes the solution to an extremely general growth equation. Moreover we will show that the usual fitness function of the TNM is simply the first order expansion of a general fitness function, and the extension of \cite{Arthur:2017} is essentially to consider all possible two species interaction terms. Thus any multi-species growth model close to equilibrium will look like the TNM. This section is important motivation towards justifying the use of the TNM to make broad conclusions about ecosystem dynamics, but it can be skipped by readers only interested in our model simulations and discussions of Gaia.

To start assume we have a number of species, $D$, with each species labelled $i$ having population $N_i$. Define the total population as $N = \sum_i N_i$ where this and all the sums are over all extant species. Define the species vector $\vec{N} = (N_1, N_2, \ldots, N_D)$ and the relative population vector as $\vec{n} = (n_1, n_2, \ldots, n_D)$ where $n_i = \frac{N_i}{N}$ is the relative population. The most general growth equation is simply that the rate of change of $i$'s population, $N_i$, is some function of all the other individuals in the system. 

Assuming that all members of the same species are equivalent (in particular that there is no aging) we can write down a set of general growth equations, known as the generalised Lotka-Volterra equations \cite{Campbell:1961},
\begin{equation}\label{eqn:glk}
    \frac{dN_i}{dt} = N_i f_i(\vec{n}, N)
\end{equation}
where $f_i$ is a fitness function. Any reasonable function $f_i$, meant to model an ecology should have at least one equilibrium at $\vec{N} = 0$ i.e. no individuals present. We can expand $f_i$ around this equilibrium
\begin{align}\label{eqn:generaleqn}
    \frac{dN_i}{dt} = N_i \bigg(
    f_i(\vec{0}, 0) + \sum_j \frac{df_i}{dn_j} (\vec{0}, 0) n_j + \frac{df_i}{dN} (\vec{0}, 0) N + \\ \nonumber 
    \frac{1}{2} \sum_{jk} \frac{d^2f_i}{dn_jdn_k} (\vec{0}, 0)n_j n_k + 
     \sum_j \frac{d^2f_i}{dn_jdN} (\vec{0}, 0)n_j N
    + \frac{1}{2}\frac{d^2f_i}{dN^2} (\vec{0}, 0)N^2
    +
    \ldots \bigg)
\end{align}

Each term in the sum represents a different type of interaction, for example $f_i(\vec{0}, 0)$ gives a constant growth rate independent of any other species while $\frac{df_i}{dn_j} (\vec{0}, 0) n_j$ represents the effect of species $j$ on the growth of $i$. These growth equations can even be constructed directly by considering one, two and higher point interactions  represented in diagrammatic form \cite{Baez:2012}. 

To give a simple example, equation \ref{eqn:generaleqn} reduces to the logistic growth equation for a particular choice of $f_i$. If
\begin{equation*}
    f_i = r - \mu N
\end{equation*}
for some constants $r$ and $\mu$ then
\begin{equation}
    \frac{dN_i}{dt} = N_i ( r - \mu N )
\end{equation}
For a single species $N = N_i$ and this is the usual logistic growth model with growth rate $r$ and $\mu$ the inverse carrying capacity.

Truncating the expansion at linear order gives the usual TNM fitness function, \cite{Christensen:2002}. To see this we set the constant term to zero, which is equivalent to saying only relative fitness should affect reproduction rates. We define
\begin{align*}
    \frac{df_i}{dn_j} (\vec{0}, 0) &= J_{ij}\\
    \frac{df_i}{dN} (\vec{0}, 0) &= -\mu
\end{align*}
$\mu$ is set to the same value for every species for simplicity and $J_{ij}$ is the strength of the effect of species $j$ on species $i$. With these definitions, the fitness function is

\begin{equation}
    f_i = \sum_j J_{ij} n_j - \mu N
\end{equation}

and hence the growth equation is

\begin{equation}\label{eqn:glktnm}
    \frac{dN_i}{dt} = N_i \left( \sum_j J_{ij} n_j - \mu N \right)
\end{equation}

which allows for species $j$ to affect species $i$ though the matrix element $J_{ij}$ and therefore model co-evolution.

An agent based model explicitly represents the individuals of each species and allows each agent the chance to reproduce and die. We say that the an agent based model `solves' equation \ref{eqn:glk} for some choice of fitness function if the average growth rate of each species is given by that function. If the agents of species $i$ have reproduction probability per individual $p_i$ and death probability $d_i$ then the expected growth rate of species $i$ in the agent based model is 
\begin{equation*}
    \frac{dN_i}{dt} = N_i (p_i - d_i)
\end{equation*}
The reproduction probability $p_i$ can be made a function of $f_i$, a fitness function. Thus $p_i = p(f_i)$ where $p$ is a function that maps $f_i$ into a properly normalised probability. It is usually chosen to be a sigmoid function of some sort, so that species with low fitness do not reproduce and species with high fitness do.

Expanding $p(f_i)$ to linear order around an equilibrium gives
\begin{equation*}
    p(f_i) \simeq a + bf_i.
\end{equation*}
for some constants $a$ and $b$. By re-scaling the fitness function we can cancel $d_i$ against the constant term $a$ and absorb the factor $b$ into the definition of fitness. This gives
\begin{equation*}
    \frac{dN_i}{dt}  = N_i f_i
\end{equation*}
The standard TNM fitness function is
$$
f_i = \sum_j J_{ij} n_j - \mu N 
$$

thus the TNM, an agent based model, `solves' the general growth equation, \ref{eqn:glktnm}, in the sense that its average behaviour is the same.

Seen in this light, the agent based model is more fundamental than the growth equations which describe its average behaviour. One key feature that the agent based model has, which cannot be replicated by the differential equations, is the generation of diversity, for example by mutation or immigration. This is for the simple reason that if the population of any species drops to zero, $N_i = 0$, then $\frac{dN_i}{dt} = 0$ and so $N_i$ remains at zero for all time. The only species we can study in the differential equation setting are the ones we start with.

As we will argue that it is the continual generation and storage of diversity that leads to Gaian effects, we must rely on agent based simulations. However the arguments above hopefully convince the reader that the TNM is not a contrived or toy model, designed to show some particular feature, but a rather direct way to add mutation to the very general growth model, equation \ref{eqn:glk}. Additionally, the usual form of the fitness function chosen in the TNM literature is quite generic and any fitness function has the same form close to an equilibrium. Of course in practice somewhat arbitrary decisions must be made in order to simulate the model e.g. the form of the probability function $p$ or the values of the matrix elements $J_{ij}$ but other work has demonstrated e.g. \cite{Arthur:2017b}, that the fundamental micro and macro dynamics of the model are very robust.

\section{Model Description and Macroscopic Observables}\label{sec:modeldes}

The specific version of the TNM we will use was developed in \cite{Arthur:2017} and modified here slightly. To motivate it we truncate equation \ref{eqn:generaleqn} at second order and rename some of the variables

\begin{equation}\label{eqn:tgaia}
  \frac{dN_i}{dt} = N_i \bigg(
     \sum_j J_{ij} n_j  -\mu N  \\ \nonumber 
    - \sum_j K_{ij}  N_j - \nu N^2  \bigg)
\end{equation}

We neglect 3-point interactions i.e. set $\frac{d^2f_i}{dn_jdn_k} (\vec{0}, 0)  = 0$ and define $\frac{d^2f_i}{dn_jdN} (\vec{0}, 0) = -K_{ij}$ and $\frac{1}{2} \frac{d^2f_i}{dN^2} (\vec{0}, 0) = -\nu$ and we used $n_j N = N_j$. 
The fitness function
\begin{equation}\label{eqn:fitness}
    f_i = \sum_j J_{ij} n_j- \sum_j K_{ij}  N_j - \mu N - \nu N^2
\end{equation}

is what we will use in our TNM simulations. As discussed in \cite{Arthur:2017} $K_{ij}$ can be thought of as a species-environment coupling matrix, where the matrix element represents the effect of $j$ on the environment of $i$. This fitness function is modified from the one in reference \cite{Arthur:2017} by the $-\nu N^2$ term, where $\nu$ is a small constant damping term, which is helpful to remove the occasional runaway growth scenarios described in that work and the $K_{ij} N_j$ term, which in \cite{Arthur:2017} has an extra factor of $n_j$. As we will show below, this change has no effect on the qualitative conclusions of that work.

To support our idea of Entropic Gaia we must measure biomass, diversity and habitability. In our model, biomass will be measured with $N$ the total population. Diversity, $D$, can be measured by counting the number of extant species, however previous work \cite{Becker:2014} has shown that the number of `core' species, $D_C$, is of more interest, since these drive the dynamics of the model. A core species will be identified as in that reference: any species whose population is at least 5\% of that of the most populous species is identified as a `core species'. All other species will be referred to as `cloud species'.

To motivate the definition of habitability, consider the following equation for the total population growth
\begin{align}\label{eqn:netlog}
    \frac{dN}{dt} &= \sum_i \frac{dN_i}{dt} = 
    \sum_{ij} N_i J_{ij} n_j
    -\sum_{ij} N_i K_{ij} N_j
    - \mu \sum_i N_i N
    - \nu \sum_i N_i N^2 \\ \nonumber
    \frac{dN}{dt} &= 
  N r - (\mu - E) N^2 - \nu N^3
\end{align}

Where $r = \sum_{ij} n_i J_{ij} n_j$ is the net growth rate and the habitability $E$ is defined as
\begin{equation}\label{eqn:hab}
    E = - \sum_{ij} n_i K_{ij} n_j
\end{equation}

The name habitability is justified by recognising equation \ref{eqn:netlog} with $\nu = 0$ as a logistic model with carrying capacity at equilibrium $\frac{r}{\mu - E}$. In practice $\nu$ will be set to a very small value ($\nu = 5\times10^{-6}$) so it only affects runs with extremely large populations. In a system with carrying capacity $\frac{r}{\mu - E}$ larger $E$ means greater carrying capacity which is how we define `habitability'.

Stability will be measured by counting the number of core rearrangements during some time interval. In an equilibrium the TNM has two or more mutually symbiotic species with large populations, the core. Periodically a new species arises which disrupts this symbiotic network and a new one appears. We call this a `quake' and count the number of such quakes by observing when the species that make up the core change. These core rearrangements also correspond to abrupt changes in the total population $N$.

The final macroscopic quantity of interest to us is the entropy, which has been defined in a number ways \cite{Roach:2017, Roach:2017a}. Here we use the definition based on the species abundance distribution. Let $Y_a$ be the number of species with a population of $a$ individuals. Let $L_a = E[Y_a]$ be the expected value of $Y_a$. In practice this is calculated by averaging over a large number of independent model runs. The entropy is
\begin{equation}
    S = -L \sum_a \pi_a \log \pi_a
\end{equation}
where $L = \sum_a L_a$ and $\pi_a = \frac{L_a}{L}$ is the probability for a species to have a population of $a$.

\subsection{Simulation Details}
The implementation of the TNM has been described numerous times \cite{Arthur:2017, Christensen:2002}. Thus we briefly describe our implementation and key parameters, referring the reader to other work for more detail. Species are defined by a binary number, which we call a genome, of length $L = 20$. The basic update step of the model is
\begin{enumerate}
    \item Choose an individual at random and calculate its reproduction probability, $p_i$. 
    \item With probability $p_i$, copy the individual, where there is a probability $p_{mut}$ to flip each of the `bits' in the genome.
    \item Choose a random individual and kill it with probability $p_k$.
\end{enumerate}
As usual in the TNM literature, we define a timescale of a `generation' as the time required to iterate over the basic reproduction/death loop $N/p_k$ times, where this number is recalculated at the end of each generation. 

The reproduction probability is
\begin{equation}
    p_i = \frac{1}{1+e^{A-f_i}}
\end{equation}
and
\begin{equation*}
    f_i = \sum_j J_{ij} n_j  - \mu N  - \sum_j K_{ij}  N_j - \nu N^2  
\end{equation*}

as in equation \ref{eqn:tgaia} and we set $A = 0$. Each of the elements of the inter-species coupling matrix $J$ is set to zero with probability $0.75$ (so that not all species interact directly). The non-zero elements are sampled from a probability distribution that is symmetric around zero and has infinite support, in particular, as is standard in the literature, we use a normal product distribution scaled by $C=100$. All of the off-diagonal elements of the species-environment coupling matrix $K$ are non-zero (so that every species affects the environment of all others) and are selected from the same distribution as $J$ but scaled by a factor $\sigma = 0.05$. The other parameter values used in this work are: $\mu = 0.1$, $p_{mut} = 0.01$, $p_k = 0.2$.

\section{Selection by Survival} \label{sec:sbs}

\begin{figure}
    \centering
        \includegraphics[width=\textwidth]{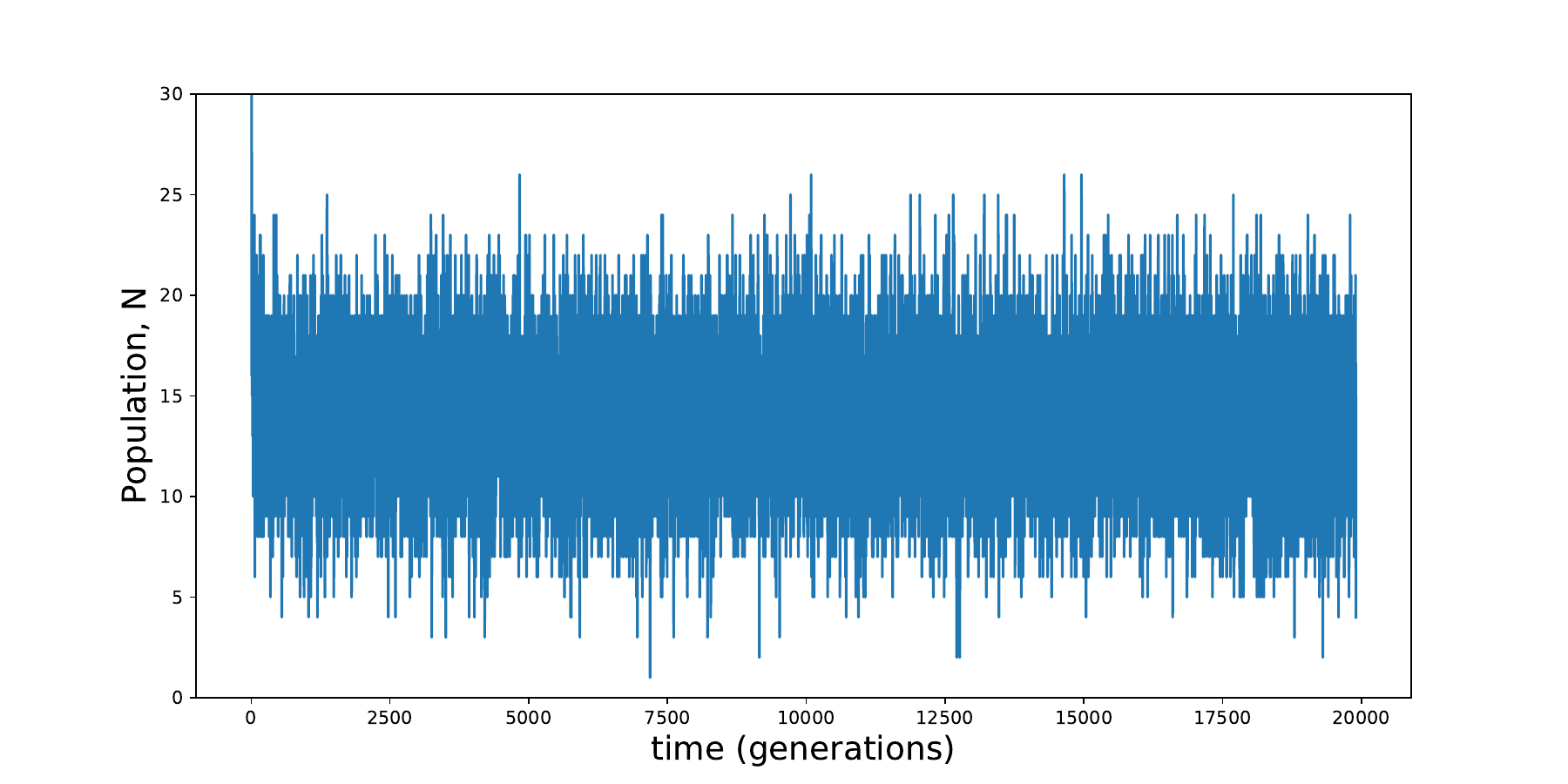}
        \includegraphics[width=\textwidth]{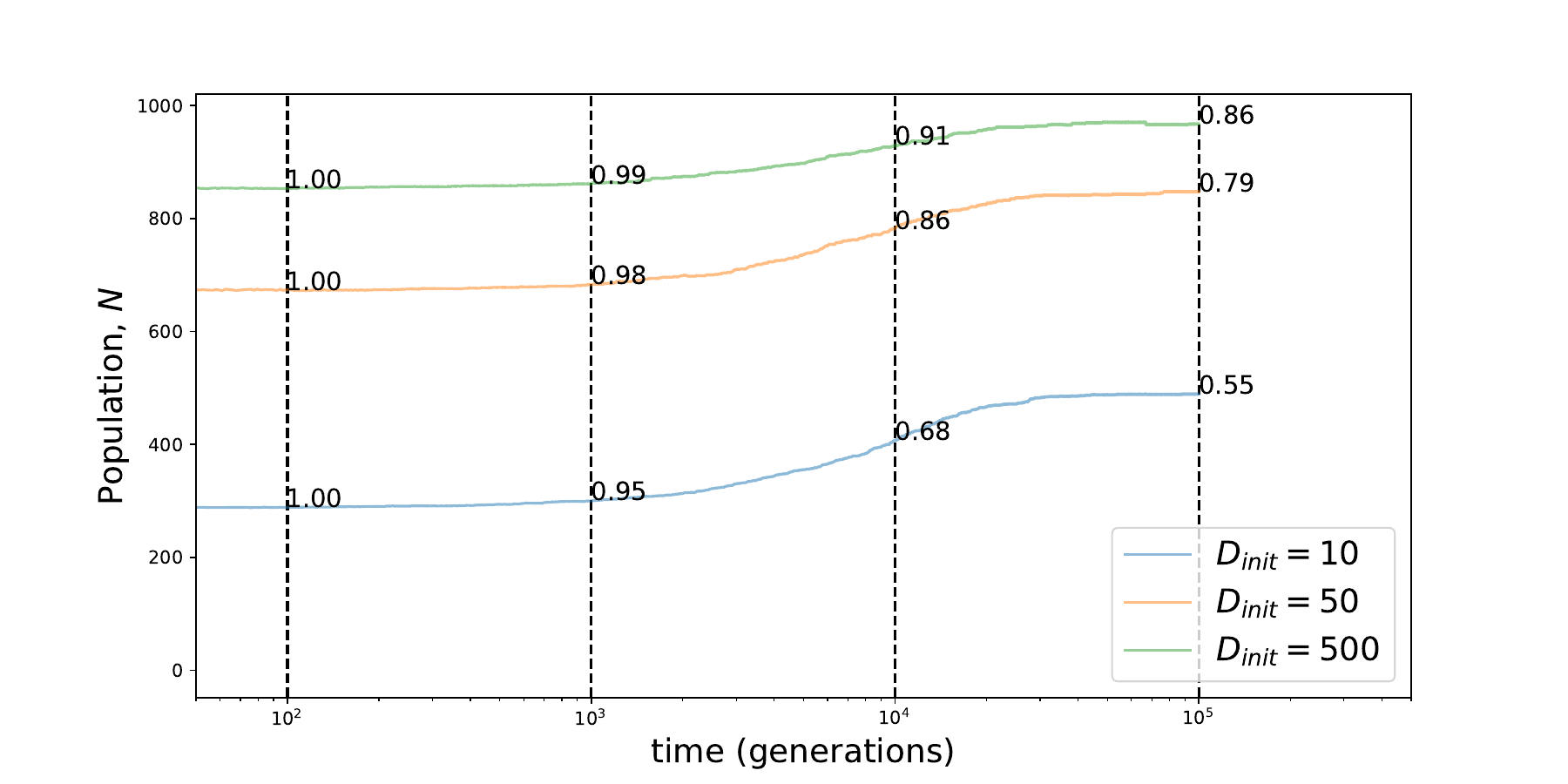}
    \caption{Top: total population as a function of time for a single realisation of the selection by survival experiment with number of species $D_{init} = 50$. The low population makes the system vulnerable to large fluctuations which can take the total population down to the fixed point $N=0$ before 100000 generations have passed. Bottom: The total population of all surviving systems for different values of $D_{init}$. The numbers show what fraction of the 1000 runs have survived to this point.}
    \label{fig:sbs}
\end{figure}

Dolittle's idea of selection by survival \cite{Doolittle:2014} can be tested directly using this framework. Previous work \cite{Nicholson:2018a} used a different model (the Flask Model \cite{Williams:2007}) to look at the survival of populations where the extinction probability was a function of the temperature. The TNM undergoes fluctuations through random birth and death processes which can lead to spontaneous extinctions if a fluctuation takes the system to zero total population, $N=0$, see Figure \ref{fig:sbs}. Thus unlike \cite{Nicholson:2018a} we don't have to add an external extinction probability to the model.

We set the mutation rate $p_{mut} = 0$, so that no new species are generated through reproduction events. We start a total population of 500 individuals, divided into $D_{init}$ different species and allow the system to evolve according to the dynamics described above. These species are Dolittle's set of ``potentially  immortal  but  non-reproducing  individuals'' \cite{Doolittle:2014}. 

\begin{figure}
    \centering
        \includegraphics[width=\textwidth]{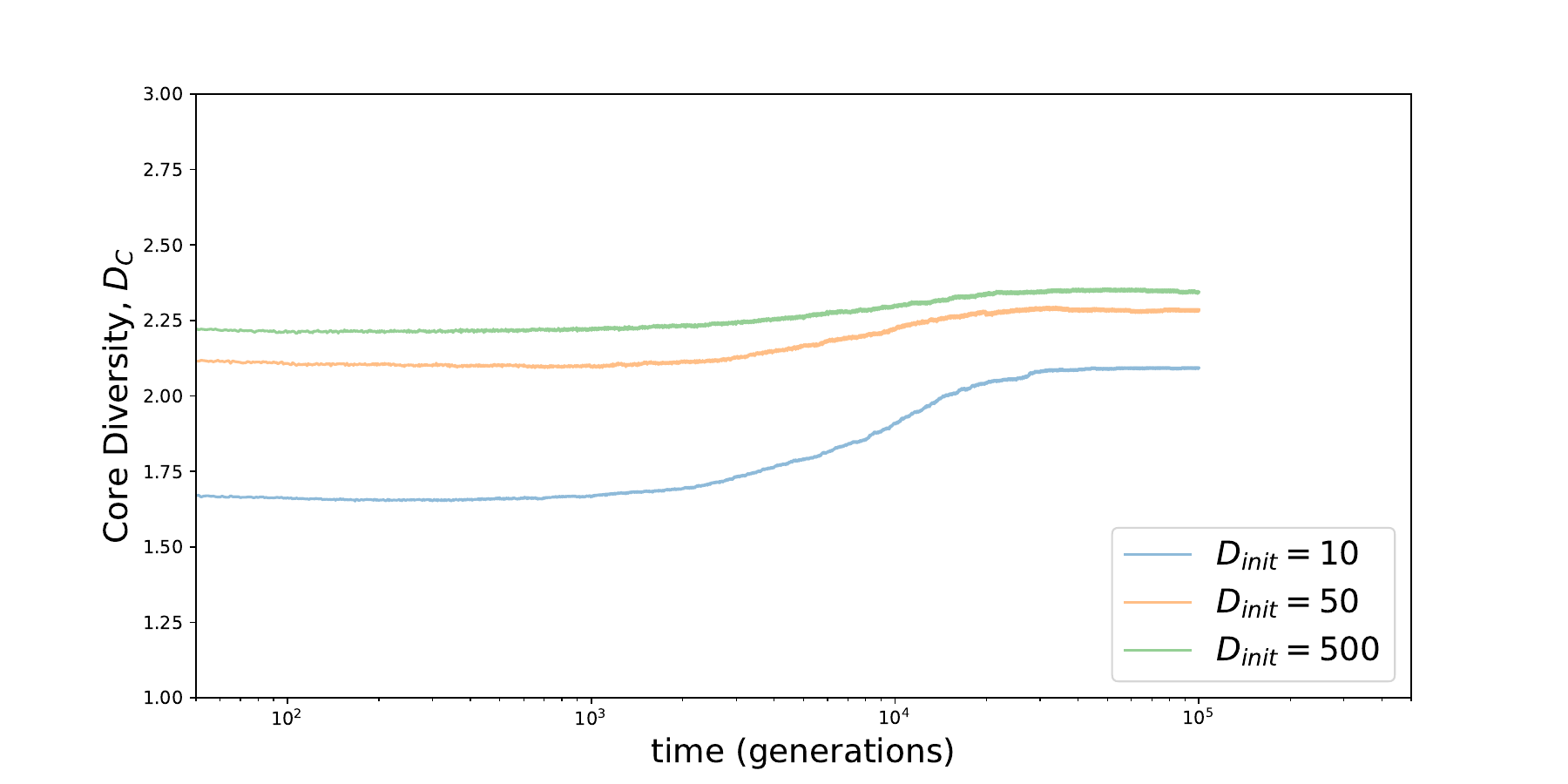}
        \includegraphics[width=\textwidth]{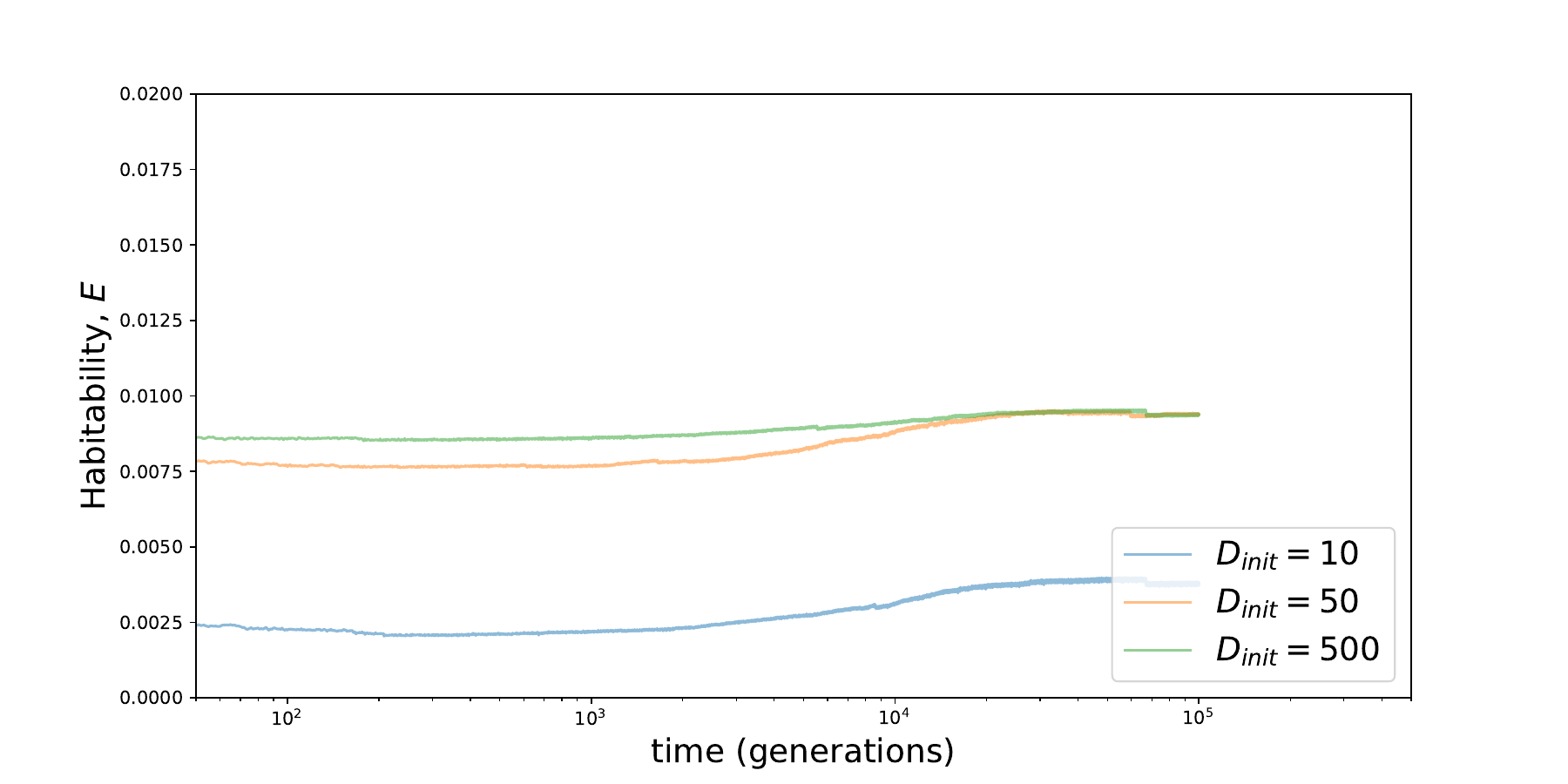}

    \caption{Top: The average number of core species $D_C$ across all surviving systems for different values of $D_{init}$. Bottom: The average habitability $E$ across all surviving systems for different values of $D_{init}$.}
    \label{fig:sbsM}
\end{figure}

The bottom of Figure \ref{fig:sbs} shows the average result of 1000 runs of this experiment for different values of $D_{init}$. We see that the average population of the surviving runs increases with time. This is simply a result of the random extinction of low population runs. Figure \ref{fig:sbsM} shows the macroscopic quantities core diversity, $D_C$, and habitability, $E$. These quantities also increase for the same reason, surviving runs tend to have higher core diversity and habitability. An interesting feature which will be important for our arguments later is that \textit{systems starting with a higher diversity of species end up systematically higher in population and habitability}.

These simple experiments show that selection by survival can lead to an apparent increase in biomass, diversity and even carrying capacity on average, for the trivial reason that at later times many of the systems with low values have gone extinct. This would seem to make Gaia a tautology - persistent systems must have some characteristics which allow them to persist. In the TNM framework these characteristics are strong interspecies couplings and positive species-environment interactions, in reality these are feedback loops and bio-geochemical cycles \cite{Lenton:2011}. Since observers must live on a persistent planet characteristics like this must be observed. This `tautological Gaia' has little explanatory power, it is essentially an anthropic argument: things are the way they are because otherwise we wouldn't be around to wonder about them. Thus selection by survival offers little guidance on some of the questions raised in the introduction on the likelihood and prevalence of life in the universe.

However there are a number of important issues with this view. This version of the model doesn't allow mutation. The generation of new species and frequent disruptions of equilibrium are a crucial aspect of TNM phenomenology as well as Earth history. Allowing for mutation enables new species to disrupt equilibria and cause the system to make large moves around the configuration space. As we will show in the Section \ref{sec:tnmsim}, this leads to increasing biomass, diversity and habitability \textit{within a single run}. This is not due to `unfit' systems dying off, but is a consequence of allowing multiple selections within the same history. This is the idea of `sequential selection' which we now discuss.

\section{Sequential Selection}\label{sec:ss}

Sequential selection as described in \cite{Nicholson:2018b, Betts:2007} is a cyclic process of quasi-stable periods where life and the environment are in some kind of homeostasis interrupted by `near fatal resets' after which a new homeostasis is established. 

\begin{figure}
    \centering
        \includegraphics[width=\textwidth]{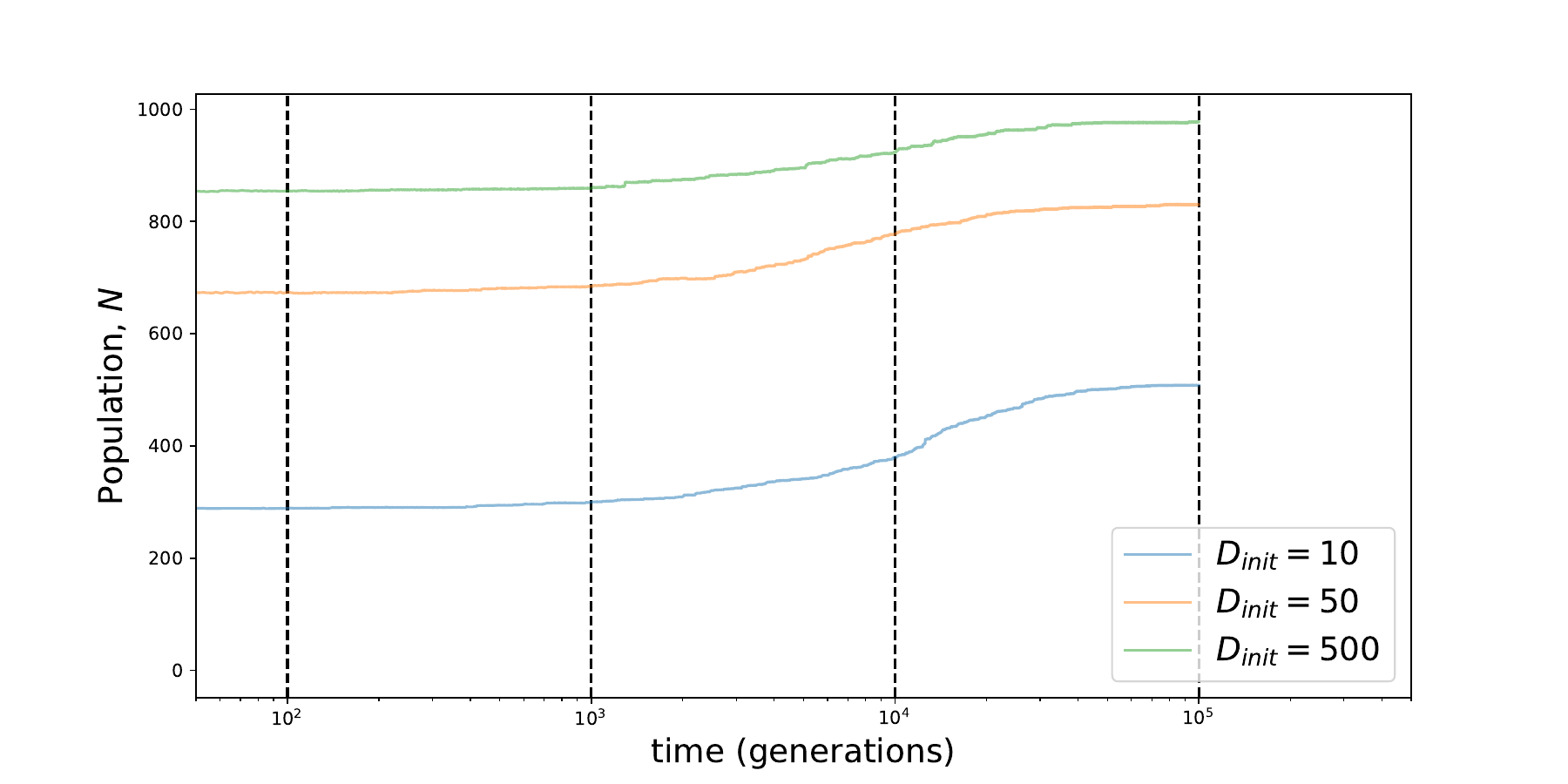}
        \includegraphics[width=\textwidth]{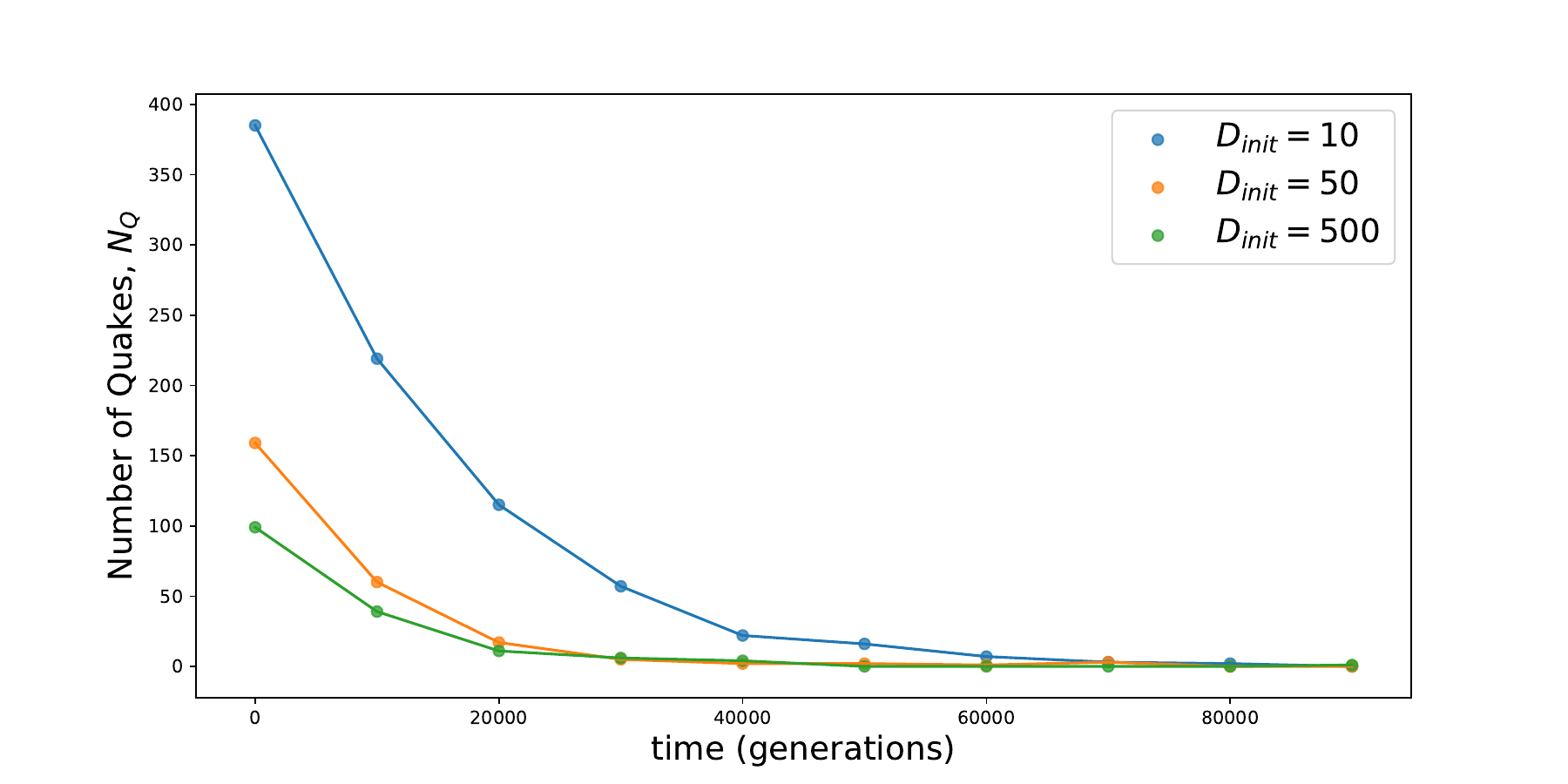}
    \caption{Top: The average population $N$ across all surviving systems for different values of $D_{init}$. Bottom: The number of resets within each $10^4$ generation windows of the $10^5$ simulation duration for different values of $D_{init}$.}
    \label{fig:ss}
\end{figure}

To simulate this we perform the same experiments as in section \ref{sec:sbs} i.e. TNM dynamics with no mutation, but when a system goes extinct, $N=0$, we restart with a new set of $D_{init}$ species. We perform 1000 runs of 100000 generations. The top plot in Figure \ref{fig:ss} shows that the population at late times is likely to be high, as is the diversity and habitability (not shown). This is for the simple reason that systems with low population are vulnerable to extinction and die out. When they do they are replaced by another system, this happens over and over until we land on a configuration which does not go extinct within the remaining time period of the experiment. 

With our choice of parameters most low population runs die out in a few thousand generations, thus are quickly weeded out and at late times all systems are likely to be in highly stable states. We can see this in the bottom plot in Figure \ref{fig:ss}, which shows the number of resets within each window of 10000 generations. The number of resets required decreases rapidly with $t$. We note once again that higher values of $D_{init}$ lead to higher final populations and a more rapid onset of stability. That is \textit{a larger starting pool of species gives rise to more stable and populous equilibria.}

These experiments show that when observing the system at a random point in its history we are most likely to land in a long lived quasi-equilibrium. Thus Gaia is once again an observer effect - long stable periods are required to generate complex organisms and leave significant fossil evidence. These periods necessarily have features conducive to stability. Similar to selection by survival Gaia occurs for tautological or anthropic reasons.

 As observed in our experiments, here and in the previous section, large populations and increased stability are associated with a bigger pool of species, larger $D_{init}$. \cite{Nicholson:2018b} claims `Overall, these studies show that sequential selection is satisficing rather than optimising, because, unlike natural selection, it cannot refine regulatory mechanisms over time.' This is certainly true, in this experiment there is a completely different regulatory mechanism after every reset. In the TNM framework this means different sets of interactions between species i.e. $J_{ij}$ and $K_{ij}$. However the dependence on $D_{init}$ shows that the regulatory mechanisms that the system ends up using are better when there are more species to choose from.

Now we finally come to the crucial issue separating the TNM with mutation from Selection by Survival and Sequential Selection. The TNM automatically performs sequential selection. A stable configuration is found which is disrupted by a quake event caused by a mutant species, after which another stable configuration is found. As with sequential selection, at later times the model is more likely to be in a persistent state, have a large population, positive species environment interactions and high diversity. However there are additional effects, related to the effect of increasing $D_{init}$, that makes these resets less random and instead tend more towards stability and Gaian behaviour.

\section{The Tangled Nature Model}\label{sec:tnmsim}
\begin{figure}
    \centering
    \includegraphics[width=\textwidth]{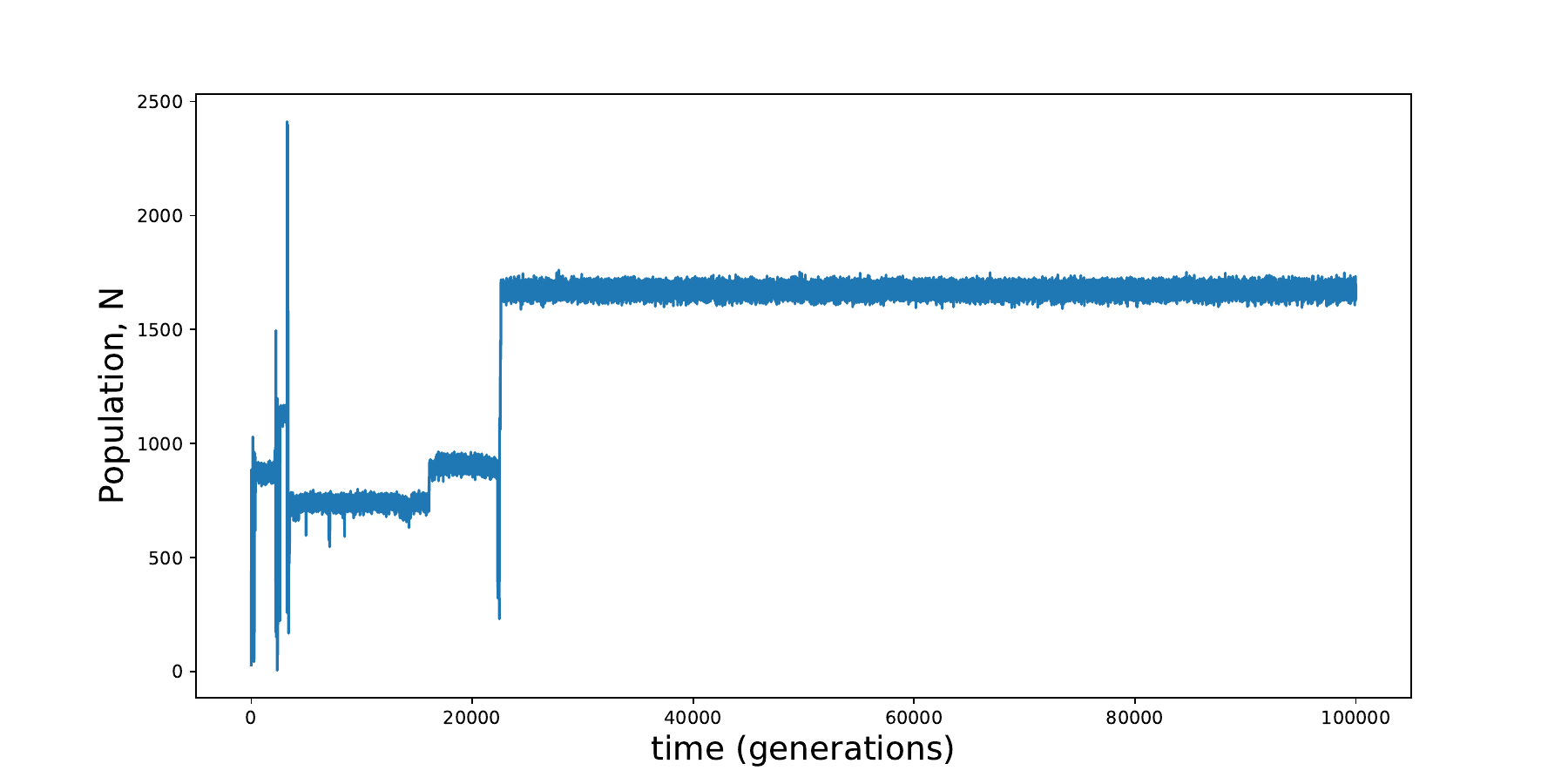}
    \includegraphics[width=\textwidth]{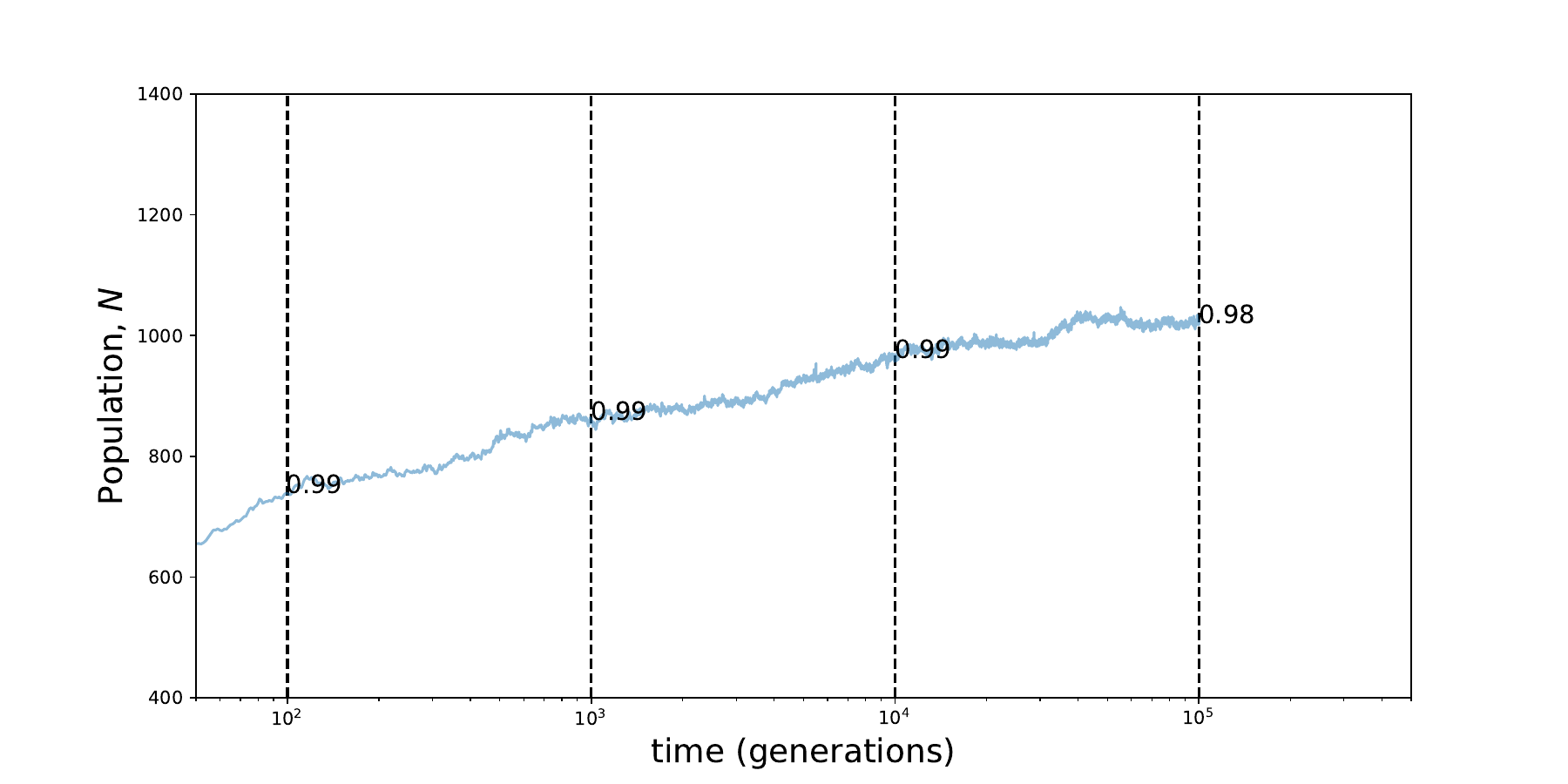}
\caption{Top: Population versus generation number for a single TNM history. Bottom: Average total population over 1000 TNM histories. The numbers indicate the proportion of surviving systems at the time indicated by the dashed line. Unlike selection by survival the majority of systems survive the entire duration of the experiment. }    
    \label{fig:tnmpop}
\end{figure}

In this section we turn mutation back on and set $p_{mut} = 0.01$. We start with 500 individuals of a single species. The top panel of Figure \ref{fig:tnmpop} shows a single TNM run, demonstrating sequential selection, the abrupt shifts in the equilibrium population. The bottom panel in Figure \ref{fig:tnmpop} shows the population versus time, averaged over a thousand runs. The total population increases logarithmically and, crucially, virtually every run survives for $10^5$ generations, demonstrating that differential survival among the different model runs is not responsible for the observed population increase.

\begin{figure}
    \centering
    \includegraphics[width=\textwidth]{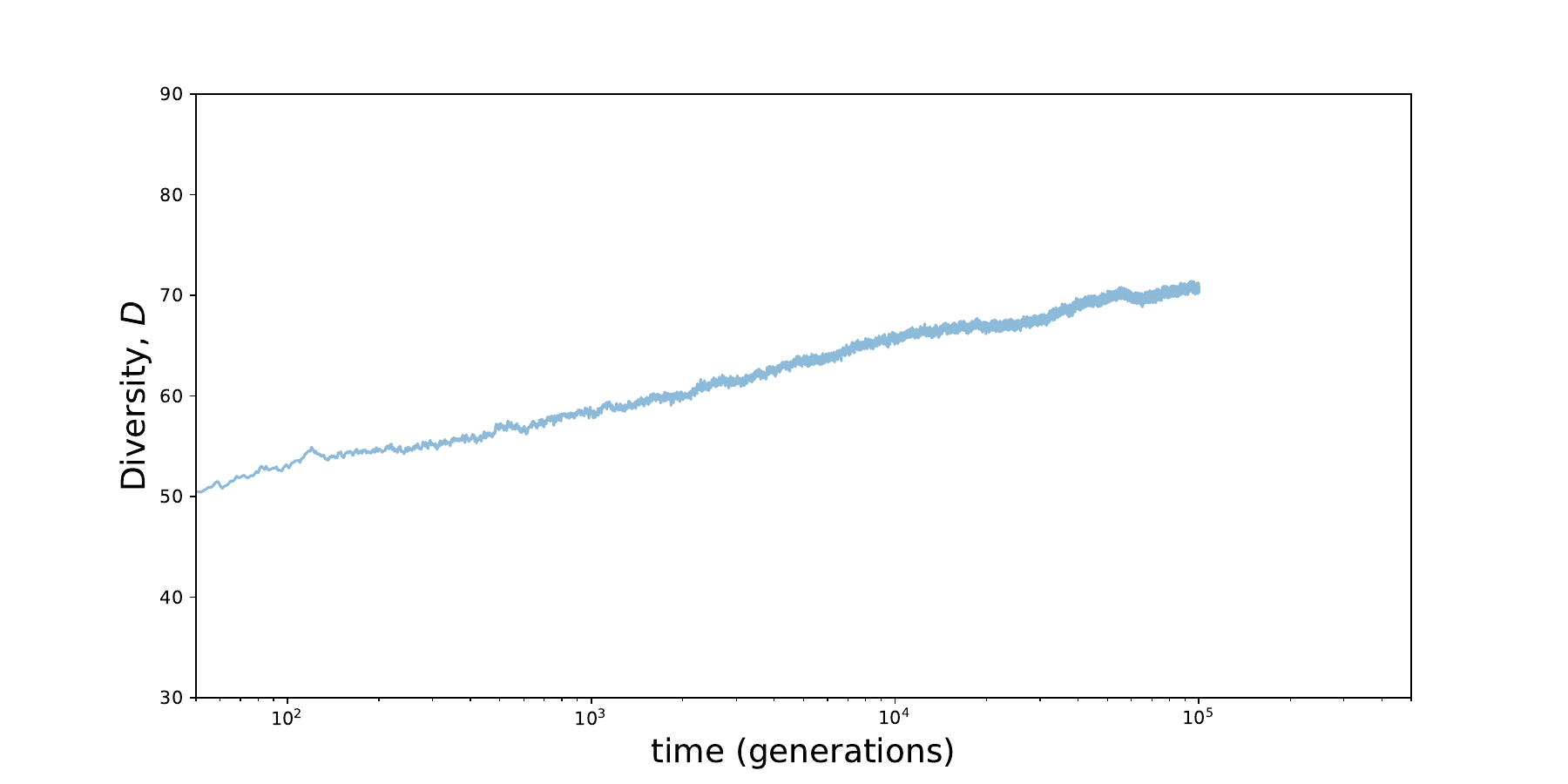}
    \includegraphics[width=\textwidth]{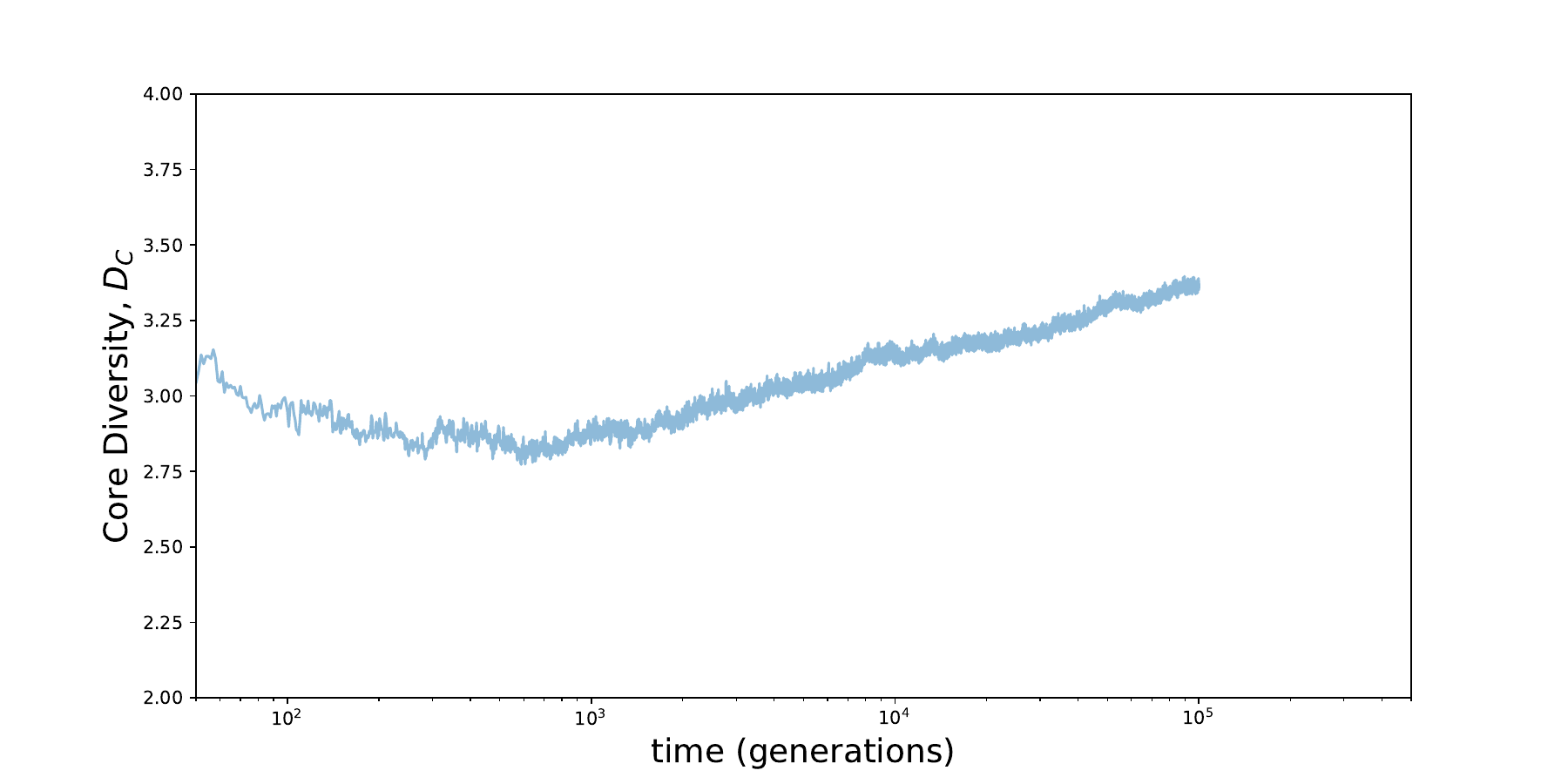}
    \caption{Top: Average total diversity, $D$, over 1000 TNM histories. Bottom: Average core diversity, $D_C$, over 1000 TNM histories. The initial high value in the first 100 generations is an artefact of the 5\% definition of a core \cite{Becker:2014} which can lead to anomalies when the total population is low.}
    \label{fig:tnmdiv}
\end{figure}

\begin{figure}
    \centering
    \includegraphics[width=\textwidth]{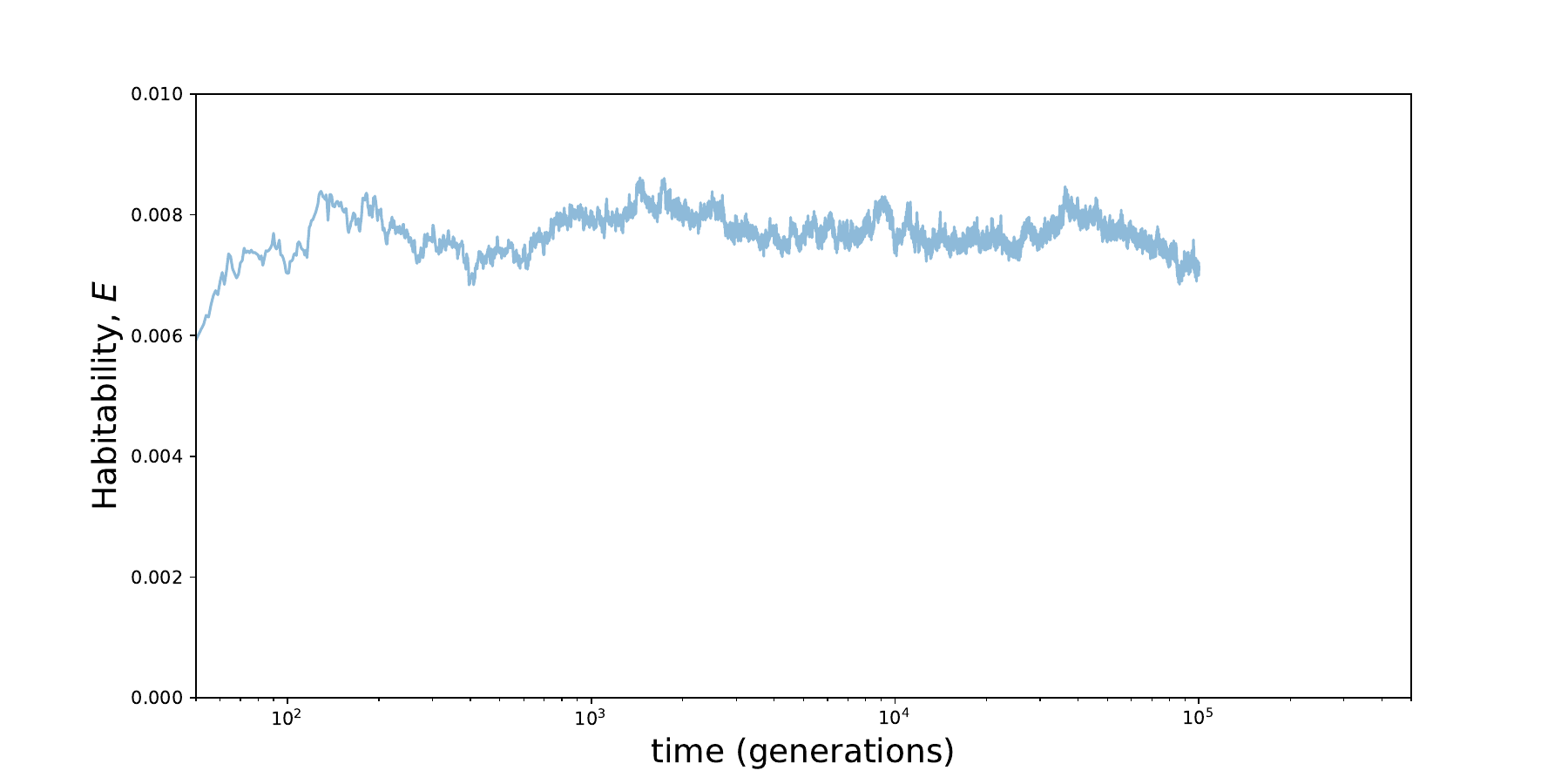}
    \includegraphics[width=\textwidth]{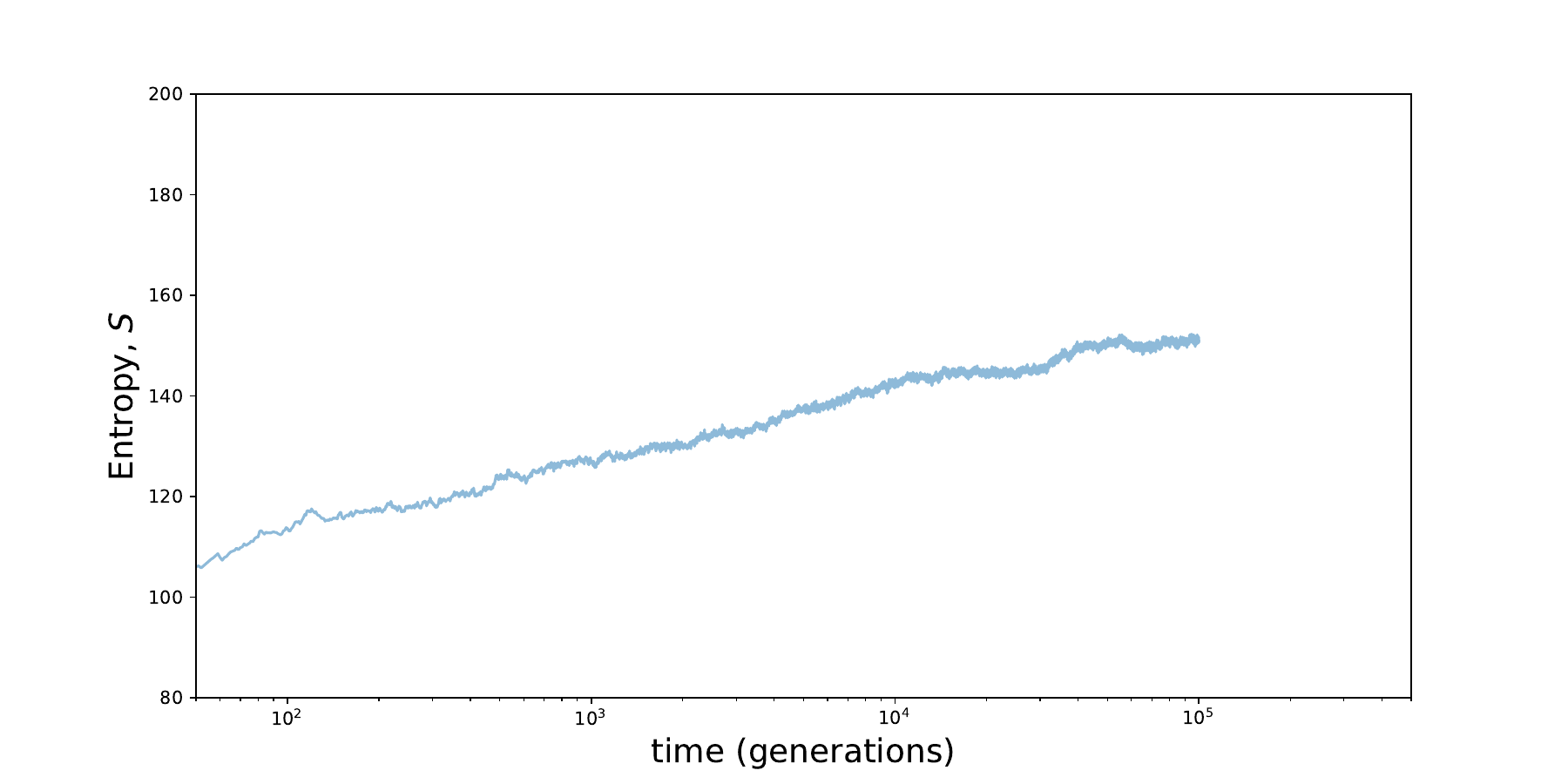}
\caption{Top: Average habitability, $E$, over 1000 TNM histories. Bottom: Average entropy, $S$, measured from the species abundance distribution accumulated over 1000 TNM histories.}
    \label{fig:tnmenthab}
\end{figure}

Figures \ref{fig:tnmpop} and \ref{fig:tnmdiv} show that \textit{on average} the population, diversity and number of core species all increase logarithmically with time. The top part of Figure \ref{fig:tnmenthab} shows that the habitability is consistently positive, though not always increasing. This effect was discussed in \cite{Arthur:2017}, the inter-species interaction $J_{ij}$ can be much larger than the species environment interaction $K_{ij}$. The species environment interaction is also scaled by absolute population $N_j$ - thus during a quake when all populations are quite low the effect of $K_{ij}$ is quite small, as it should be, a small population of a new species is unlikely to have an environmental impact until it grows to a significant size. However the species-environment interactions are consistently positive, showing that the species which persist at late time tend to have environment improving interactions. The bottom part of Figure \ref{fig:tnmenthab} shows the entropy increasing. Most of the entropy or disorder in the TNM is carried by the cloud \cite{Becker:2014}. As the population increases, the constant mutation rate results in a larger and more diverse cloud leading to an increase in entropy. 

\begin{figure}
    \centering
    \includegraphics[width=\textwidth]{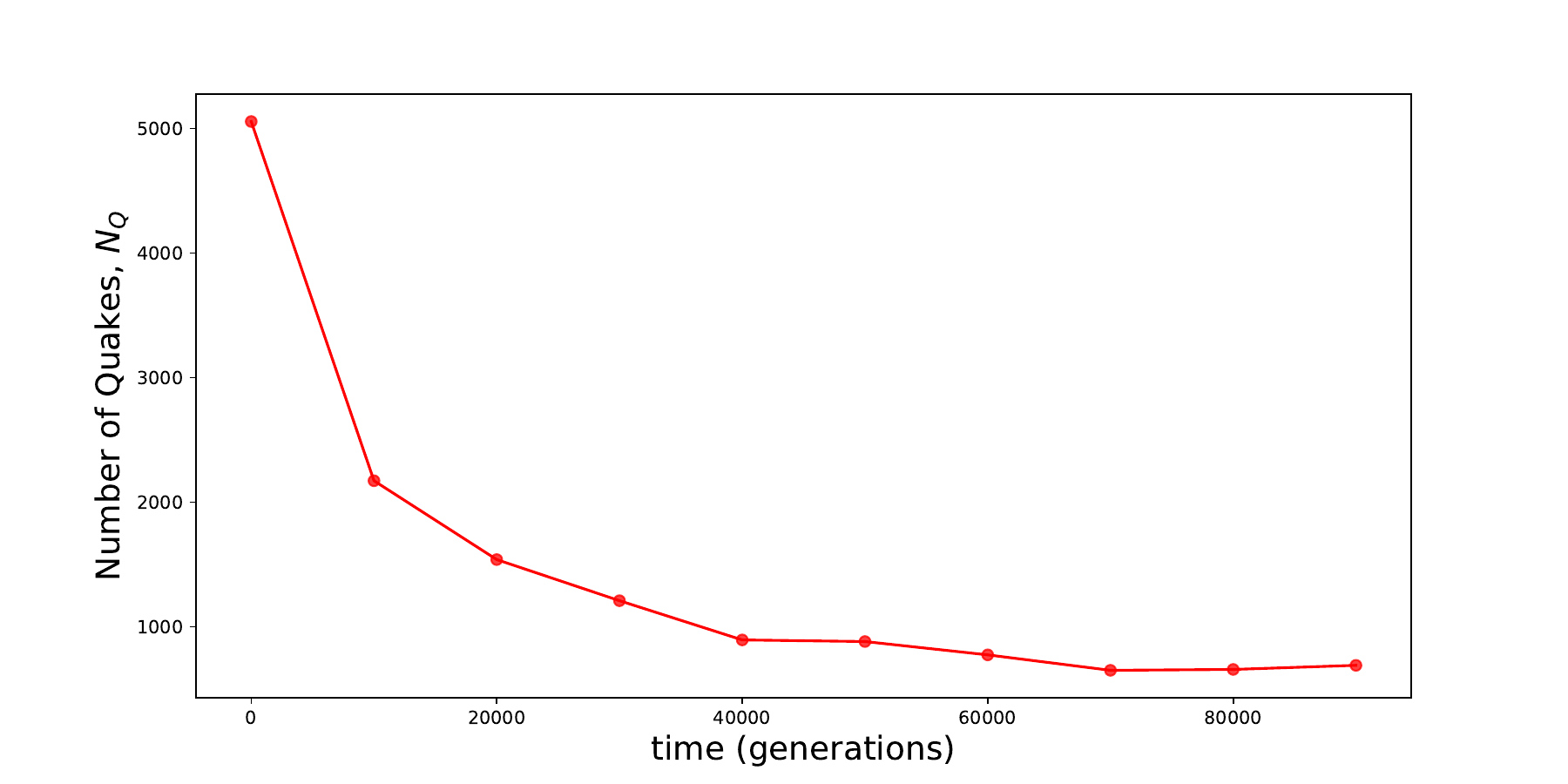}
    \includegraphics[width=\textwidth]{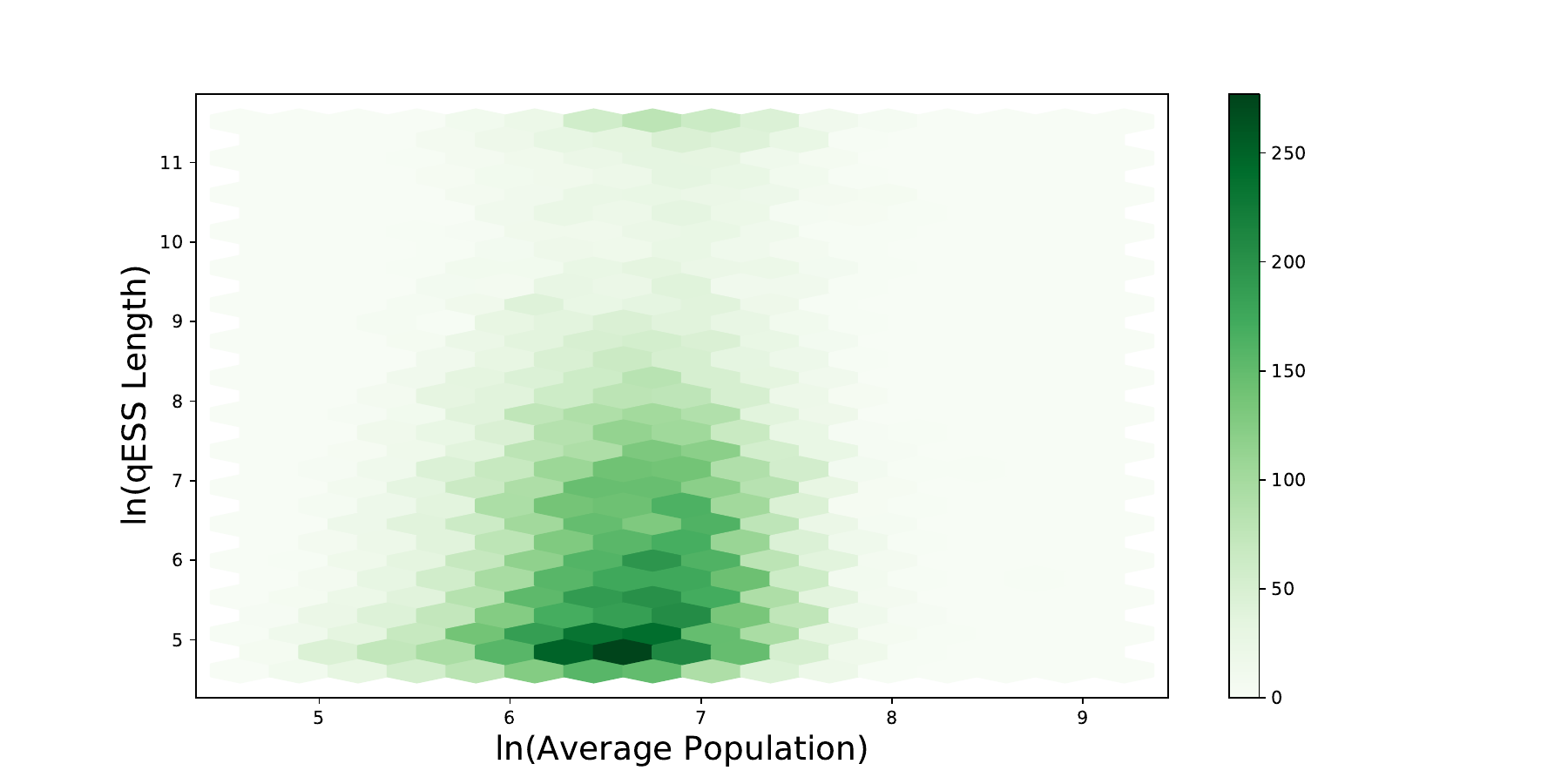}    
\caption{Top: Number of quakes observed in each 10000 generation window over 1000 TNM histories. Bottom: Density plot showing duration of stable periods versus their population, the colour bar shows the counts. The association between the (logged) population and duration is positive with Pearson's correlation coefficient 0.144 p-value $<10^{-12}$ }
    \label{fig:tnmquakes}
\end{figure}

Finally, the top panel of Figure \ref{fig:tnmquakes} shows that the number of quakes decreases on average and the systems become progressively more stable. The bottom panel shows that higher populations are positively associated with longer stable periods. The absolute counts are not necessarily significant, many more short stable periods can be observed during a run than can long ones of $O(10^5)$ generations, but the correlation between average population and time until a quake is positive and significant. This will be examined in more detail in the next section, but see also \cite{Becker:2014}.

\subsection{Understanding the TNM}\label{sec:tnmdiscussion}

Looking at the macroscopic equation, \ref{eqn:netlog}, in a quasi-equilibrium there will be a balance between the positive terms arising from inter-species symbiosis and environment improving interactions ($r$ and $E$) negative terms representing the limits to growth ($\mu$ and $\nu$). A mutant that is able to replicate rapidly will destabilise this quasi-equilibrium. Roughly, this happens if $p_i \geq p_k$ which implies
\begin{align*}
f_i \geq A - \log\left( \frac{1}{p_k} -1 \right) = f_{min}
\end{align*}
where this equation defines $ f_{min}$. Using the definition of the TNM fitness, equation \ref{eqn:tgaia}, this implies the fitness of a destabilising mutant is
\begin{align}\label{eqn:mutantfitness}
    f_i = \sum_j J_{ij} n_j  - \mu N  - \sum_j K_{ij}  N_j - \nu N^2  \geq f_{min} \\ \nonumber
   \implies \sum_j J_{ij} n_j  - \sum_j K_{ij}  N_j \geq f_{min} + \mu N + \nu N^2
\end{align}

The larger the total population $N$, the higher the barrier on the right hand side. This means strong the symbiotic interactions ($J_{ij}$) and/or environment improving interactions ($K_{ij}$) of the new species with the extant species, especially the core, are necessary. 

If such a mutant occurs it will grow exponentially according to equation \ref{eqn:glk}. This growth is at the expense of the core species, since there is a finite carrying capacity. The core species reduce in population, in turn reducing the population of the disrupting mutant species which rely on interactions with the core. This leaves the system in a state where no species makes up a significant fraction of the population. From those species still remaining - that is remnants of the core, disrupter, cloud and nearby mutants - a new core arises. The core that comes to dominate is the one which grows fastest and uses up the carrying capacity. This selects for cores with strong positive interactions with each other $J_{ij}$ and environment improving effects $K_{ij}$, since these are the ones which grow fastest according to the growth equation, \ref{eqn:glk}.

\begin{figure}
    \centering
    \includegraphics[width=\textwidth]{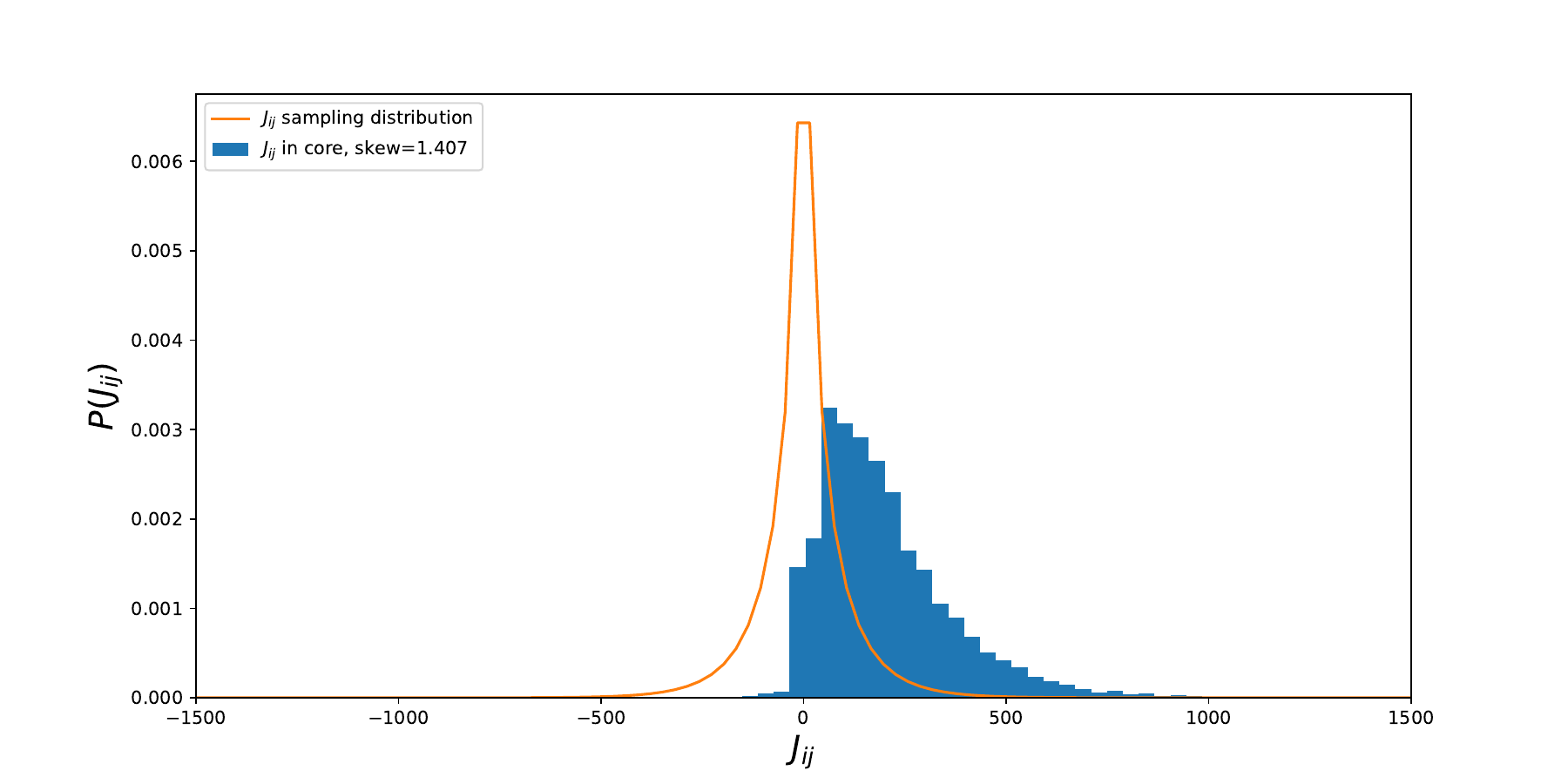}
    \includegraphics[width=\textwidth]{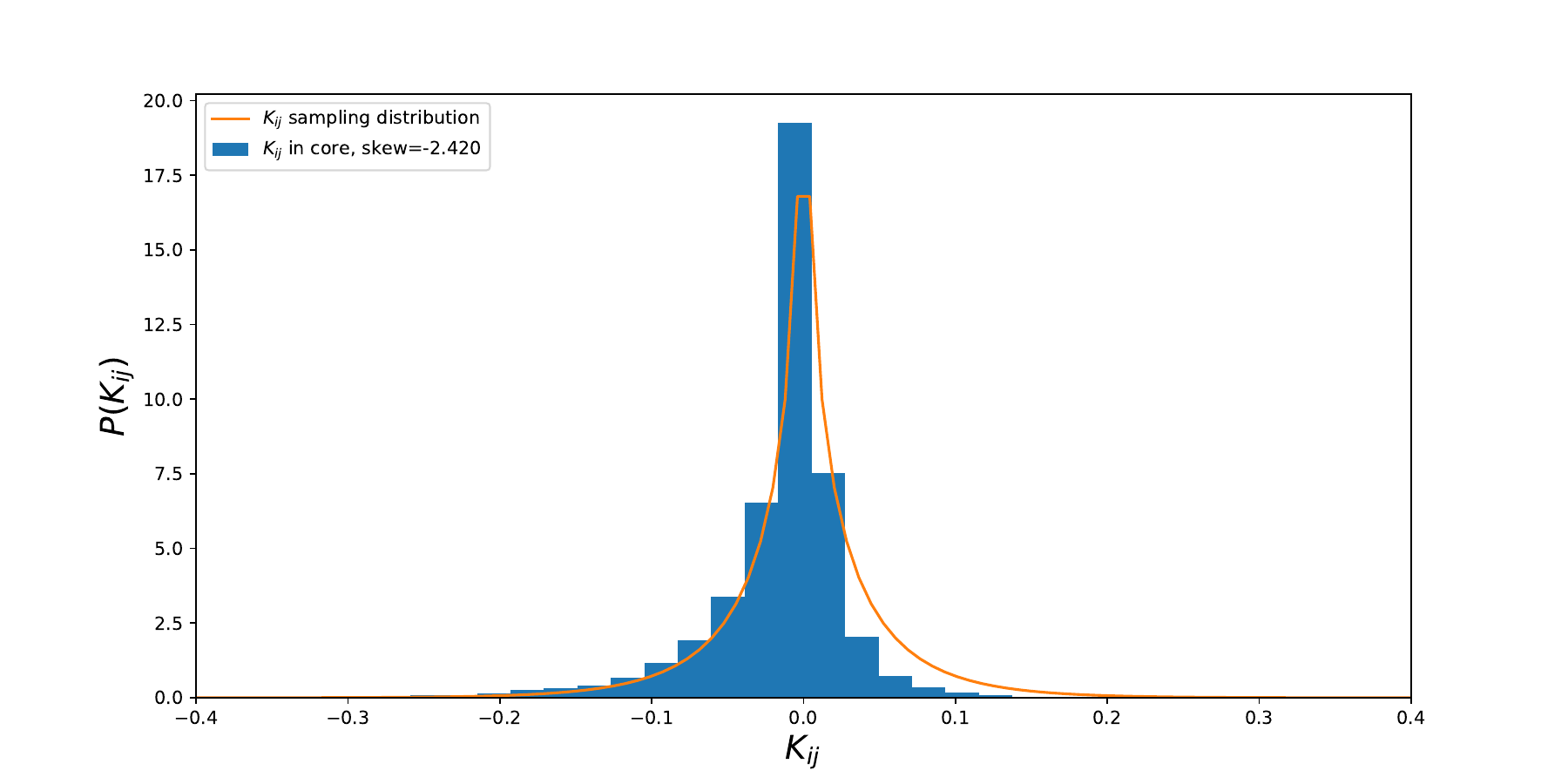}    
\caption{Top: Distribution of $J_{ij}$ values realised in core-core interactions at $t = 10^5$ generations. The distribution is significantly right skewed (z-score $41$, vanishing p-value), showing that mutually beneficial core-core interactions are selected. Bottom: Distribution of $K_{ij}$ values realised in core-core interactions at $t = 10^5$ generations. The distribution is significantly left skewed (z-score $-56$, vanishing p-value), showing that environment improving core-core interactions are selected. The degree of the skew depends on the strength of the species environment coupling, $\sigma$, e.g. compare to \cite{Arthur:2017} where stronger species environment coupling was used. }
    \label{fig:tnmhist}
\end{figure}

This can be seen in the plot of average population, Figure \ref{fig:tnmpop}, and habitability, Figure \ref{fig:tnmenthab}, since larger values of population and habitability reflect positive and negative values of $J_{ij}$ and $K_{ij}$, respectively. We can see this directly in Figure \ref{fig:tnmhist}, which shows the distribution of $J_{ij}$ and $K_{ij}$ values realised in the core after $10^5$ generations (see also \cite{Arthur:2017}). Compared to the random sampling distributions, the TNM dynamics realise highly skewed distributions and therefore leads to, on average, a beneficial biotic environment (positive $J_{ij}$) and a beneficial abiotic environment (negative $K_{ij}$). Unlike many Gaian models, the TNM agents do not explicitly regulate through feedback loops to control the environment. This is instead modelled through the $K_{ij}$ term (the effect of j on the carrying capacity of the system for species i). These figures demonstrate that the model dynamics lead to higher carrying capacities on average (also seen the plot of habitability, $E$, Figure \ref{fig:tnmenthab}). This effect, of life enhancing the conditions for life, is distinctly Gaian.

This selection effect leads to higher final populations which, from equation \ref{eqn:mutantfitness}, makes a higher barrier for the next destabilising mutant to cross. As shown in Figure \ref{fig:tnmpop} this also leads to higher core and cloud diversity over time. The higher diversity means that after a quake there is a bigger pool of potential cores to choose from, and hence the new core is more likely to be populous, diverse, habitable and stable. This is a kind of `ratcheting' sequential selection, where after every reset the number of potential new cores is larger and Gaian ecosystems are more likely to occur.

\section{Understanding Entropic Hierarchy}\label{sec:tnment}

 We discussed in \ref{sec:fl} that there is no competition between ecosystems in the modelling framework we used here, the models runs are completely independent. Thus, unlike the single organism landscape \cite{Kauffman:1989} there is no competition forcing ecosystems to move ‘uphill’ on the ecosystem fitness landscape, equation 1. The main driver of ecosystem change in the Tangled Nature Model is entropy maximisation - in other words the system will be in the most probable state \cite{Becker:2014}. This means the ecosystem will most likely be found in the area of configuration space with the longest stable periods, which is the idea of selection by survival and sequential selection discussed in sections \ref{sec:sbs} and \ref{sec:ss}.

\begin{figure}
    \centering
\includegraphics[width=0.8\textwidth]{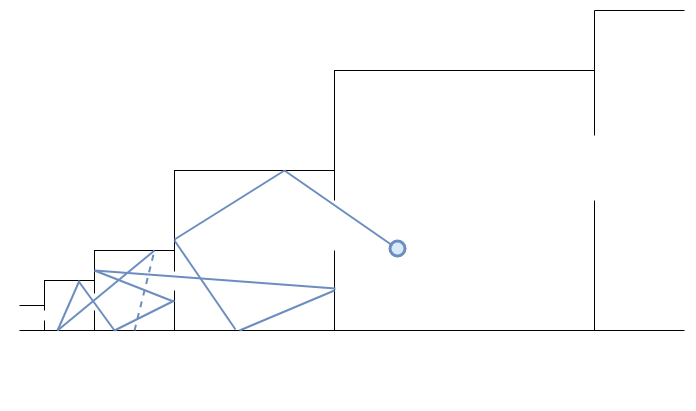}
    \caption{Another entropic hierarchy, a particle in a series of connected boxes.  In a small box there are few possible locations for the particle (low entropy) and it will quickly find an exit. In a larger box their are many possible locations (high entropy) so it will take longer to leave. Since small boxes are easier to escape, the particle is likely to be found on the right, though moving right takes longer and longer at each level.}
    \label{fig:example}
\end{figure}

In section \ref{sec:tnmsim} we argued that there is another effect occurring, which we called ratcheting sequential selection and in other contexts has been called an entropic hierarchy \cite{Arthur:2017, Barettin:2011}. To introduce this potentially unfamiliar idea consider the toy model depicted in Figure \ref{fig:example}. A particle is rebounding in a series of connected boxes that increase in size.  The collisions are assumed to be perfectly elastic, so there is no energy loss. Even if there are no forces acting on the particle it still tends to move to the right. This is because a random search will quickly find the exit to a small box, but take much longer in a large one. Entropic forces are well known in physics, for example in Brownian motion \cite{Roos:2014} and elasticity \cite{Neumann:1977}. In this example there is an apparent `force' pushing the particle to the right which we can attribute to entropy maximisation. This is to be contrasted with the situation of `pure' sequential selection, which would have the box sizes randomised. In that case the particle would tend to be in a large box but there is no tendency for it to move in either direction. 

\begin{figure}
    \centering
\includegraphics[width=0.47\textwidth]{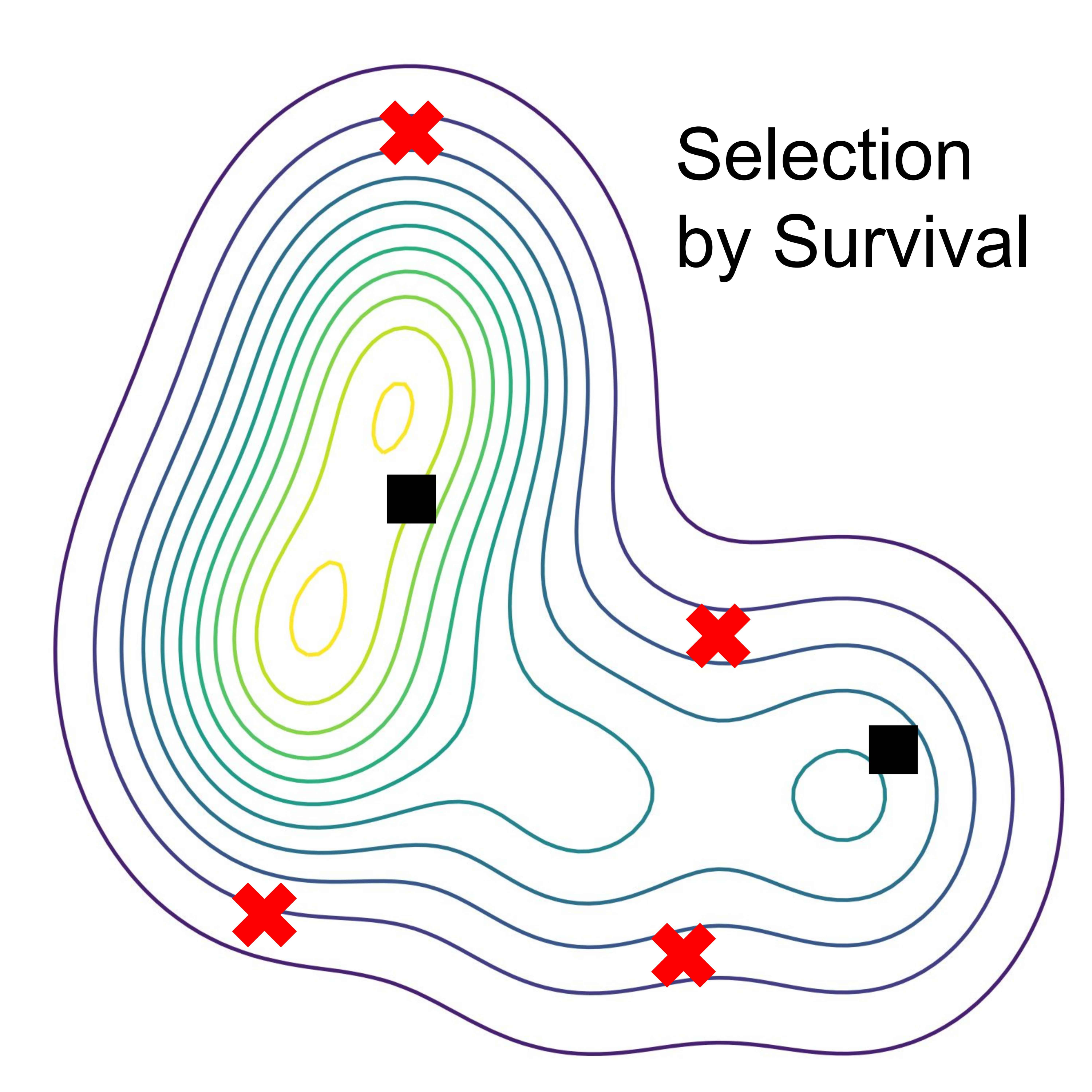}
\includegraphics[width=0.47\textwidth]{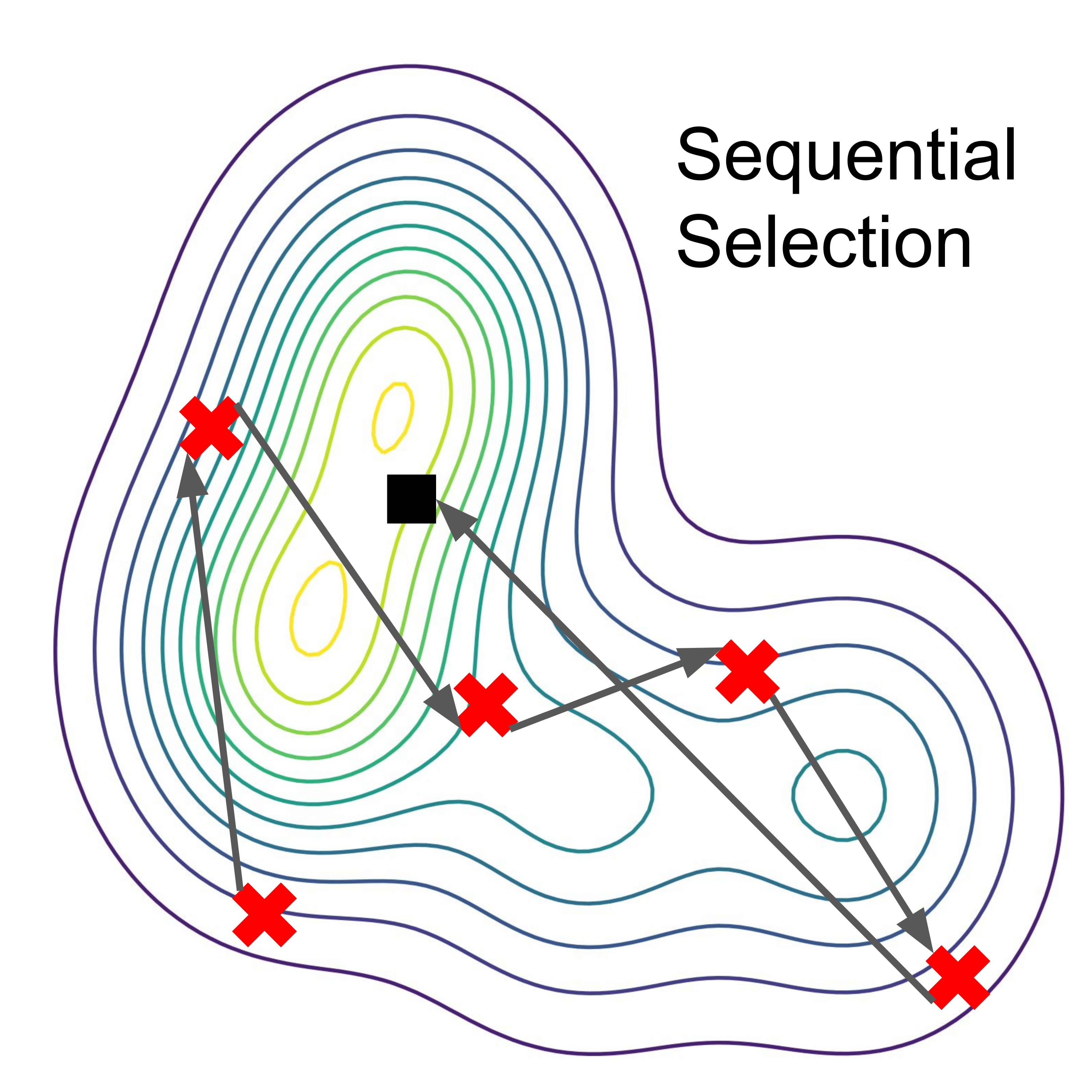}
\includegraphics[width=0.47\textwidth]{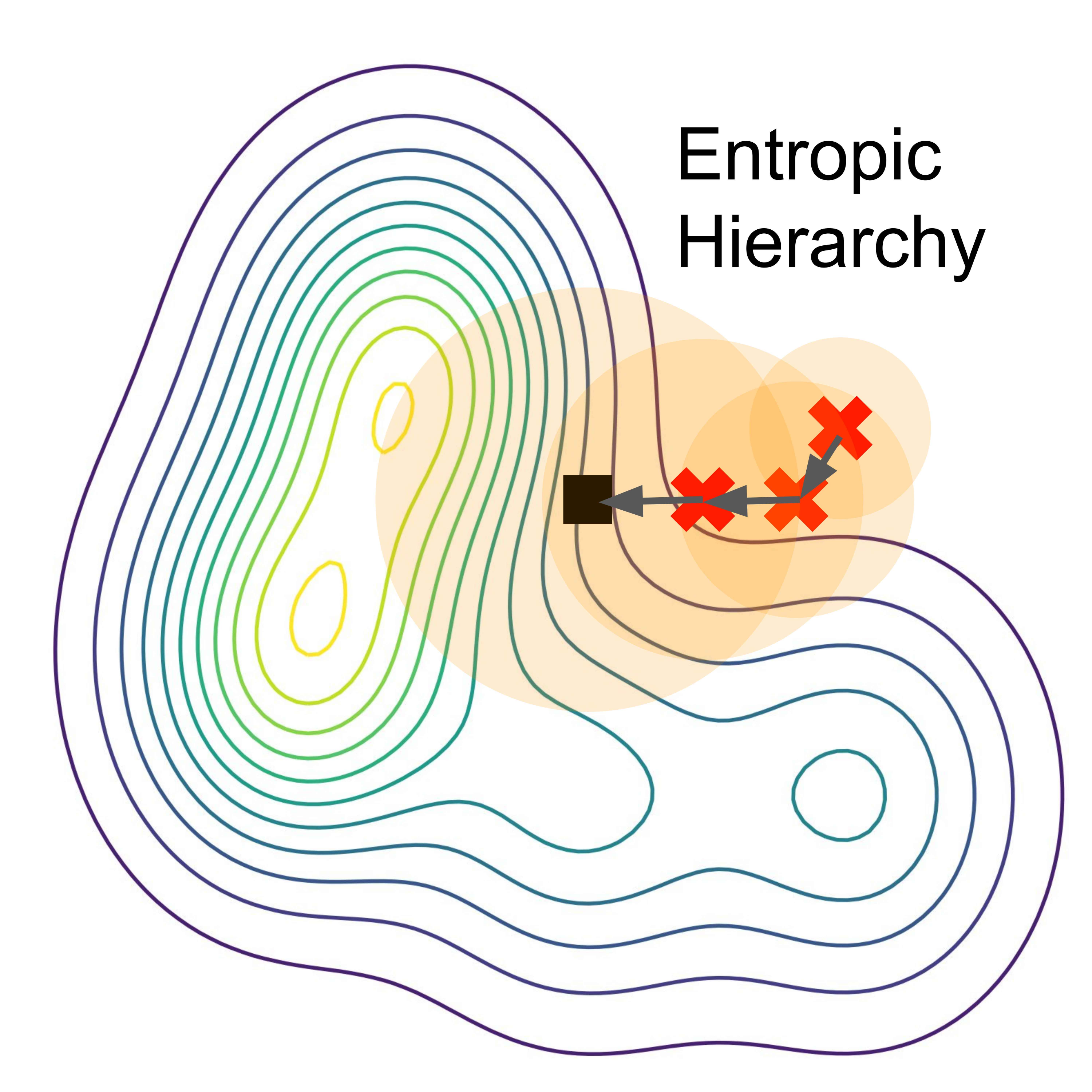}
    \caption{A representation of configuration space for the three mechanisms discussed in this paper. The red `X' indicates an ecosystem that has gone extinct, the black square represents a system which has persisted until the time of observation. }
    \label{fig:contour}
\end{figure}

Figure \ref{fig:contour} is another visual metaphor, this time the total fitness function $F$ is represented by a contour map. The three cases of selection by survival, sequential selection and entropic hierarchy are illustrated. Selection by survival randomly scatters systems across the fitness landscape \cite{Doolittle:2014}. Sequential selection moves to a random point in the space after every extinction event \cite{Nicholson:2018b}. In contrast the TNM can explore a region of the space (represented by the orange circle) during each stable period. After an extinction event the new system arises from somewhere in this region. Systems which grow quickly are likely to form the new core as discussed in section \ref{sec:tnmsim} and \cite{Arthur:2017, Becker:2014,Arthur:2017b} with the result that the new system tends to be `uphill' from the previous one , leading to the growth in population observed in Figure \ref{fig:tnmpop} and discussed in Section \ref{sec:tnmdiscussion}. 

To link these metaphors back to the TNM, consider the situation when a quake happens and a quasi-equilibrium ends. The number of possibilities from which to choose the next species network tends, on average, to be larger than after the last quake event, as demonstrated by the logarithmically increasing population and diversity shown in Figures \ref{fig:tnmpop} and \ref{fig:tnmdiv}. The system which arises to dominate and fill the vacuum has to have strong symbiotic inter-species interactions \cite{Becker:2014} and no deleterious effects on the environment \cite{Arthur:2017} so that it can use up the carrying capacity before any other potential core configuration has the chance, implying that a configuration with a large total fitness $F$ will dominate, according to the basic growth mechanism of the TNM, equation \ref{eqn:glk}, and lead to a large final population according to the same equation. 
 
Larger populations in the TNM generate more diversity, as seen in Figure \ref{fig:tnmdiv}, since each reproduction event has a constant probability to generate mutants, so more reproduction means more mutants. We have seen directly in Figure \ref{fig:sbs} that a higher starting diversity leads to higher final populations. An analogous situation exists during a quake after a core collapse, where a bigger starting pool of species leads on average to a higher equilibrium population. When the next quake happens, the same process can unfold again, this time with an even larger pool of species from which the new core can arise, which tends to result in an even larger equilibrium population, leading to the increasing average population observed in Figure \ref{fig:tnmpop}. This is how the TNM walks up the entropic hierarchy, and how the hill climbing occurs without any competition between ecosystems. 

Of course this is a stochastic process, so the system will occasionally move downhill, e.g. the single history shown in Figure \ref{fig:tnmpop}. Due to the increasing entropic barriers, equation \ref{eqn:mutantfitness}, the time between steps grows, shown in Figure \ref{fig:tnmquakes} and discussed in Section \ref{sec:tnmdiscussion} and reference \cite{Becker:2014}, meaning this is a very slow process.  Overall however, in the TNM, there is a slow but significant tendency towards greater population, diversity, life-enhancing interactions with the abiotic environment and more stability, as demonstrated in Figures \ref{fig:tnmdiv}, \ref{fig:tnmenthab} and \ref{fig:tnmquakes}. This process, the statistical tendency of life to enhance the capacity of the system to support life, is what we call Entropic Gaia.

\section{Diversity Reservoirs}\label{sec:divres}

Our results suggest that the following conditions are necessary for Entropic Gaia:
\begin{enumerate}
    \item Darwinian selection (diversity, inheritance and competition).
    \item Periodic `resets'.
    \item A reservoir of diversity.
\end{enumerate}
The first point is established by the contrast between our model experiments in section \ref{sec:sbs} with those in section \ref{sec:tnmsim}. Number 2 is a prerequisite of sequential selection \cite{Nicholson:2018b}.  In the TNM the periodic resets are generated by the model dynamics, in the same way that many of Earth's mass extinctions are suggested to have arisen from the action of life itself \cite{Lenton:2011}. However these resets could also be exogenous, e.g. an asteroid impact or volcanic activity. Point 3 is established by a comparison between pure sequential selection discussed in section \ref{sec:ss} and the full model and its analysis performed in \ref{sec:tnmsim} and \ref{sec:tnment}. Pure sequential selection is memoryless, while the the TNM builds on previous quasi-equilibria, with a more diverse starting point after each reset, leading to larger equilibrium populations (as can be observed directly in Figure 1). As argued in section \ref{sec:tnmsim} and \ref{sec:tnment}, it is the ability to build from the previous equilibrium that is the key to climbing the entropic hierarchy. 

This section will suggest three mechanisms by which the Earth system preserves diversity: seed banks, refugia and lateral gene transfer. Hopefully these examples make the idea of Gaia's memory more plausible and demonstrate how the entropic hierarchy could work in practice.

\subsection{Seed Banks}\label{sec:seedbank}

The importance of seed banks for plant ecology is well known, where seed banks help ecosystems recover from disturbances \cite{Kalamees:2002,Luzuriaga:2005} and maintain species richness \cite{Myers:2009}. Recent work has extended the concept from plants to a \textit{microbial seed bank}, defined as `a reservoir of dormant individuals that can potentially be resuscitated in the future under different environmental conditions' \cite{Lennon:2011}.

This microbial seed bank is maintained by the mechanism of dormancy \cite{Jones:2010,Lennon:2011}. Dormancy is a reversible state or life stage where metabolic activity is drastically reduced. The exact mechanisms of dormancy vary greatly \cite{Guppy:1999,Hoehler:2013,Bodegom:2007}, for example spores or condia in fungi, akinetes in cyanobacteria or bacterial cysts. Dormancy is generally observed under conditions of low resource availability, high residence time, frequent environmental perturbations and low predation \cite{Lennon:2011}. Dormant micro-organisms generally resuscitate once more favourable environmental conditions (for them) occur.


Dormancy helps maintain many properties of the microbial seed bank \cite{Lennon:2011}. For example by repeated transitions in and out of the seed bank in response to environmental changes, dormancy helps maintain high levels of microbial diversity \cite{Jones:2010}. Dormancy reduces the probability of local extinction and increases colonization success (by allowing new species to wait for favourable conditions in their new environments). 
Generally, the stability of ecosystem processes is attributed to high microbial diversity \cite{Bell:2005,Balvanera:2006,Allison:2008}. This is because high diversity leads to functional redundancy \cite{Naeem:1998}, the ability of different species to fulfill the same roles in an ecosystem. Dormancy contributes to the maintenance of this redundancy by allowing species which would otherwise compete to co-exist, with one in a dormant state. If the competing species goes locally extinct the dormant species can re-emerge (e.g. in response to increased resource availability) and fill the same metabolic role as the previous species, maintaining stability in the ecosystem. 

From the Gaian perspective the microbial seed bank provides a biological mechanism for how high diversity can be maintained in the face of perturbation. The ability of dormant microbes to survive in extreme environments and remain viable over very long periods is precisely what is needed to maintain or restart Gaian regulation after massive ecological shocks e.g. asteroid impacts or snowball earths \cite{Lenton:2011}. The microbial seed bank is thus a useful reservoir of diversity.

\subsection{Climate Refugia}\label{sec:refugia}

The Earth's climate is not uniform. This simple fact means that species (microbes or more complex organisms) can retreat to more hospitable regions in the face of unfavourable climatic changes or advance into more hospitable regions as climate improves. Bennet and Provan \cite{Bennett:2008} identify numerous types of \textit{refugia}. The classical refugia refer to small regions where species of flora \cite{Bennett:1991} and fauna \cite{Kotlik:2006,Sommer:2006} survived glaciations, though with greatly reduced numbers. 
Cryptic refugia are favourable microclimates or low population densities in the inhospitable region, so that species do not go completely extinct in these areas \cite{Stewart:2001}. 

Isolated local refugia are sometimes refereed to as \textit{microrefugia} \cite{Rull:2009,Dobrowski:2011} to distinguish e.g. the Iberian peninsula as a potential refugium during northern glaciation from e.g. a favourable micro-climate in a sheltered coastal area. We can also have the inverse: what is hostile for one species may be ideal for another, so as one species expands from its refugium another may contract. For example we currently observe altitudinal migrations of species, some going uphill, some going down \cite{Blake:2000}.

Ashcroft \cite{Ashcroft:2010} identifies that there is a somewhat loose definition of refugia in the literature e.g. the distinction between micro and macro-refugia. However for our purposes it suffices to recognise that these pockets of diversity are suggested to have existed by palaeontological and biogeographic data \cite{Provan:2008} as well as genetic data \cite{Petit:2003}. For Gaia these refugia act as physical stores of genetic information and allow rapid re-expansion after large scale perturbations.

\subsection{LGT and Genomic Space}\label{sec:lgt}

Goldenfeld and Woese \cite{Goldenfeld:2007} refer to microbes as `gene-swapping collectives'. This is in reference to the ubiquity of lateral gene transfer (LGT) between microbes \cite{Ochman:2000}. According to this viewpoint we should not view microbes as distinct species with individual characteristics but as collectives that are swapping and discarding genetic material in response to environmental pressures. LGT limits our ability to fight disease \cite{Darmon:2014} but gives bacteria and archaea a tremendous ability to rapidly adapt to different environments.

Aminov \cite{Aminov:2011} gives examples of LGT occuring in soils, aquatic ecosystems, animal guts and within biofilms. Interestingly the rate of LGT increases substantially in response to stress, e.g. UV light \cite{Claverys:2006}. Thus LGT is likely to be key in responding to perturbations of Gaia, as a means of enabling microbes to rapidly adapt to changing conditions. LGT does not result in constant addition to the microbial genome. As microbes are under tremendous selection pressure they will rapidly lose non-functional genes \cite{Mira:2001}. LGT is a mechanism by which adaptive mutations can rapidly spread. Together with the microbial seed bank containing a large reservoir of genes with various metabolic functions, LGT can enable microbes to spread environmental adaptations once they are `learned' by the system. In this way useful genes can be transferred between species, so that even if the originating species goes extinct its genes live on in the extant bacterial population, keeping the pool of useful genes higher than it would be otherwise.

\section{Summary and Conclusion}\label{sec:conc}

The TNM is a simplified model, but very general. A number of other models find links between increasing stability and diversity (see the review of \cite{Nicholson:2018b} especially \cite{Dyke:2013}). We have shown in section \ref{sec:tnm} that a generic multi-species model of population growth gives, essentially, the TNM in the lowest order of approximation. This implies that TNM-like dynamics might be present across a wide array of systems. 

Using this framework we have discussed three paradigms which could lead to Gaia: Selection by Survival, Sequential Selection and Entropic Hierarchy. Selection by Survival and Sequential Selection lead to anthropic reasoning - Gaia is observed because Gaia is necessary for there to be observers. Our proposal is for Entropic Gaia. This is the idea that co-evolutionary systems increase in entropy \cite{Roach:2017, Roach:2017a}, and that increasing entropy is associated with a number of `Gaian' features, including higher total biomass, higher diversity, reduced rate of endogenously generated extinctions and the positive effect of life on the abiotic environment. These features are observed in the TNM in section \ref{sec:tnmsim} and further discussed in section \ref{sec:tnment}. Given the generality of the TNM and the mechanisms for maintaining diversity discussed in section \ref{sec:divres} we postulate Entropic Gaia as a way to generate Gaia, without resorting to anthropic reasoning.

Anthropic principles have little if any predictive power, Entropic Gaia on the other hand makes a number of predictions. For example it predicts we would observe a `punctuated equilibrium' in the fossil record as stable periods are disrupted  \cite{Gould:1972,Bak:1993}. The tendency towards stability and diversity should also be reflected in the fossil record and indeed \cite{Newman:1999} reports a decline in the extinction rate and an increase in diversity observed over time. The authors of \cite{Sibani:2021} describe how these hierarchical systems, and specifically the TNM, can be described in the framework of `record dynamics'. This framework makes a number of concrete predictions e.g. log-Poisson quake statistics, which can be tested against the fossil record. Unfortunately, this interesting comparison is beyond the scope of this paper.

We have discussed how seed banks and refugia provide a plausible biological mechanism for maintaining a reservoir of species diversity, while LGT provides a mechanism where even individual genes can be stored. High levels of ambient diversity allow recovery after major quakes and higher diversity makes it easier for persistence enhancing feedback loops to arise.

We believe this model - sequential selection combined with `memory' can provide a plausible explanation for why Gaia is observed on Earth and why Earth's history has unfolded as it has - where useful innovations seem to have seldom been lost but rather built upon e.g. photosynthesis required the earlier evolution of at least two types of anoxygenic photosynthesis reaction centres and pigment synthesis pathways \cite{Canfield:2006}, see also the idea of `evolutionary learning' in \cite{Watson:2016}. These innovations then provide the means to realise the bio-geophysical feedback loops by which Gaia maintains herself.

These ideas also have a number of implications for our search for life on other planets and the so-called Fermi Paradox \cite{Goldblatt:2015, Lenardic:2016}. If   Entropic Gaia  is correct, the longer life has persisted on a planet the longer it is likely to keep persisting and growing in diversity and abundance, so older inhabited planets would be more likely to host diverse and complex life.

\section*{List of Symbols}

LK-model variables:
\begin{itemize}
    \item $i$: Species label/genotype
    \item $N_i$: Number of individuals of type $i$
    \item $D$: Diversity, number of distinct species.
    \item $N$: Total population = $\sum_{i} N_i$
    \item $f_i$: Fitness of species $i$
    \item $F$: Ecosystem fitness = $\sum_i N_i f_i$
    \item $n_i = N_i/N$ : Relative population.
    \item $p_i$ : Reproduction probability for species $i$.
    \item $d_i$ : Death probability for species $i$.    
\end{itemize}
Terms in fitness function:    
\begin{itemize}
    \item $\mu$ : Inverse carrying capacity.
    \item $J_{ij}$ : Inter-species interaction matrix.
    \item $K_{ij}$ : Species-environment interaction matrix.
    \item $\nu$ : Damping factor.    
\end{itemize}
TNM parameters:
\begin{itemize}
    \item $L$ : Genome length.
    \item $p_{mut}$ : Mutation rate.
    \item $p_{k}$ : Death rate
    \item $A$ : Fitness function offset (controls reproduction rate of individuals with zero fitness).
    \item $C$ : Scaling factor for $J$.
    \item $\sigma$ : Scaling factor for $K$.
\end{itemize}
Macroscopic Observables:
\begin{itemize}
    \item $r$: Net growth rate.
    \item $D_C$ : Number of core species.
    \item $N_C$ : Number of individuals in the core.
    \item $E$ : Habitability.
    \item $S$ : Entropy.    
    \item $N_Q$ : Number of quakes.
\end{itemize}

\section*{Code}
The code used in this work is available at \url{https://github.com/rudyarthur/Tangled-Nature}.

\section*{Acknowledgements}
The authors would like to thank David Wilkinson for some very helpful suggestions. This work was partly funded by the Leverhulme Trust through a research project grant RPG-2020-82.

\end{document}